\documentclass[journal]{IEEEtran}
\usepackage{amsmath,amsthm,amssymb}
\usepackage{graphicx}
\usepackage{flushend}
\usepackage{subcaption}
\usepackage{array}
\usepackage{float}
\usepackage[ruled,vlined,linesnumbered]{algorithm2e}
\usepackage{setspace}
\usepackage{tikz}
\usepackage{hyperref}
\usepackage{multirow}
\usepackage{multicol}
\usepackage{blkarray}
\allowdisplaybreaks

\let\oldnl\nl
\newcommand{\nonl}{\renewcommand{\nl}{\let\nl\oldnl}}

\theoremstyle{plain}

\newtheorem{theorem}{Theorem}

\newtheorem{corollary}{Corollary}
\newtheorem{example}{Example}

\theoremstyle{remark}

\DeclareMathOperator{\sgn}{sign}

\DeclareMathOperator{\F}{\mathbb F}
\DeclareMathOperator{\W}{\mathbf W}

\DeclareMathOperator{\bS}{\mathbf S}

\renewcommand{\ell}{l}

\newcommand{\set}[1]{\left\{{#1}\right\}}

\newcommand{\mD}{\mathcal{D}}

\newcommand{\mF}{\mathcal{F}}

\newcommand{\mR}{\mathcal{R}}

\newcommand{\mZ}{\mathcal{Z}}

\newcommand{\bF}{\mathbb{F}}

\begin{document}

\title{Window Processing of Binary Polarization Kernels}
\author{Grigorii Trofimiuk, ~\IEEEmembership{Student Member,~IEEE,} Peter Trifonov, ~\IEEEmembership{Member,~IEEE}
\thanks{G. Trofimiuk, and P. Trifonov are with ITMO University, Saint-Petersburg, Russia.
 E-mail: \{gtrofimiuk,pvtrifonov\}@itmo.ru
 
 This work was supported by Government of Russian Federation (grant 08-08).
 
 This work was partially presented at the Information Theory Workshop'2018 and International Symposium on Information Theory'2019.
 
 The source code is available at 
 
 \url{https://github.com/gtrofimiuk/SCLKernelDecoder}.}}

\date{\today}
\maketitle


\markboth{IEEE Transactions on Communications,~Vol.~XX, No.~Y, July ~20YY}
{Trofimiuk, Trifonov: Window Processing of Binary Polarization Kernels}

\begin{abstract}

%

A decoding algorithm for polar (sub)codes with binary $2^t\times 2^t$ polarization kernels is presented. It is based on the window processing (WP) method, which exploits the linear relationship of the polarization kernels and the Arikan matrix.
This relationship enables one to compute the kernel input symbols probabilities
by computing the probabilities of several paths in Arikan successive cancellation (SC) decoder.

In this paper we propose an improved version of WP, which has significantly lower arithmetic complexity and operates in log-likelihood ratios (LLRs) domain. The algorithm identifies and reuses common subexpressions arising in computation of Arikan SC path scores.

The proposed algorithm is applied to kernels of size 16 and 32 with improved polarization properties.
It enables polar (sub)codes with the considered kernels to simultaneously provide better performance
and lower decoding complexity compared with polar (sub)codes with Arikan kernel.

\end{abstract}
\begin{IEEEkeywords}
Polar codes, polarization kernels, fast decoding.
\end{IEEEkeywords}
\IEEEpeerreviewmaketitle 
\sloppy
\section{Introduction}

Polar codes are a novel class of error-correcting codes, which achieves the symmetric capacity of a binary-input discrete memoryless channel  $W$. They have low complexity construction, encoding, and decoding algorithms \cite{arikan2009channel}. 
However, the performance of polar codes of practical length is quite poor.
The reasons for this are their low minimum distance and the existence of imperfectly polarized bit subchannels, which are used for transmission of data.  This causes the SC algorithm to be highly suboptimal. These problems can be alleviated by employing the successive cancellation list (SCL) algorithm  \cite{tal2015list}, as well as improved code constructions, such as polar subcodes and polar codes with CRC 
\cite{trifonov2016polar,trifonov2017randomized,wang2016paritycheckconcatenated}.   
  
Polarization is a general phenomenon and is not restricted to Arikan matrix \cite{korada2010polar}.  One can replace it by a larger matrix, called \textit{polarization kernel}, which has better polarization properties obtaining thus better finite length performance. Polar codes with large kernels were shown to provide an asymptotically optimal scaling exponent \cite{fazeli2018binary}. 
Various polarization kernels together with simplified processing algorithms were recently proposed, including $l$-formula method for medium length (up to 16) kernels \cite{huang2018simplified}, BCH  kernels with reduced trellis processing complexity \cite{moskovskaya2020design}, approximate processing based on Box and Match algorithm \cite{miloslavskaya2014sequentialBCH}, and convolutional polar kernels \cite{morozov2020convolutional}. 
However, the complexity of the existing processing algorithms for the kernels with improved polarization properties (compared with Arikan kernel)
is still too high for practical use. 


Window processing algorithm was introduced in \cite{trifonov2014binary} and applied to Reed-Solomon kernels. Independently, WP algorithm was suggested for processing of permuted Arikan matrix \cite{buzaglo2017permuted}. WP\ exploits the relationship of the considered kernels and the Arikan matrix, which enables one to compute the kernel input symbols probabilities
by computing the probabilities of several paths in Arikan SC decoding. In this paper we propose an improved version of WP, which has significantly lower arithmetic complexity and operates in LLRs domain. The proposed algorithm identifies and reuses common subexpressions arising in recursive computation of Arikan SC path scores. Further complexity reduction is achieved by exploiting some properties of maximization of these  scores.

The proposed approach allows very simple processing for some specially constructed kernels.  We discuss the application of the proposed algorithm to polar codes with $16 \times 16$ and $32 \times 32$ kernels with improved polarization properties. Namely, we consider two $16 \times 16$ kernels both having rate of polarization $0.51828$, but different scaling exponents: $3.346$ and $3.45$. We present also a $32 \times 32$ kernel with rate of polarization $0.521936$ and scaling exponent $3.41706$. For comparison, Arikan matrix has rate of polarization $0.5$ (the optimal value is 1) and scaling exponent $3.627$ (the optimal value is 2).

The proposed algorithm can be used with SCL\ decoding. Simulation results show that polar codes with the considered large kernels provide significant performance gain compared with polar codes with Arikan kernels under decoding with the same list size. Furthermore, the proposed approach results in lower decoding complexity compared with polar codes with Arikan kernel with the same performance.

The paper is organized as follows. The background on polar
codes and polarization kernels is presented in Section
\ref{sBackground}. The window processing (WP) algorithm is introduced in Section \ref{sConventionalWindow}.  Section \ref{s:CSE} presents a common subexpression identification and reusing method, which provides significant complexity reduction of WP. Some further improvements of WP are discussed in Section \ref{sWindSimpl}. The processing algorithms for the considered kernels are described in Section \ref{sEfficientProc}.  Simulation results are presented in Section \ref{sNumberic}.

\section{Background}
\label{sBackground}

In this section we give a brief introduction to polarization kernels, including polarization properties, and polar codes. We also define the problem of kernel processing and describe the processing of Arikan matrix in the approximate LLR domain.

\subsection{Notations}
For a positive integer $n$, we denote by $[n]$ the set of $n$ integers $\{0,1,\dots ,n-1\}$. The vector $u_a^b$ is a subvector $(u_a,u_{a+1},\dots,u_b)$ of some vector $u$. For vectors $a$ and $b$ we denote their concatenation by $a.b$. $M[i]$ is an $i$-th row of the matrix $M$. By $M[i,j]$ we denote $j$-th element of $M[i]$. 

\subsection{Polarizing transformation}
\label{ss:pt}
Polarization kernel $K$ is an $l\times l$ invertible matrix, which is not upper triangular under any column permutation \cite{korada2010polar}. 
Consider a binary-input memoryless channel  with transition probabilities $W(y|c), c\in \F_2, y\in \mathcal Y$, where $ \mathcal Y$ is the output alphabet. 
 An $(n = l^m, k)$ polar code is a linear block code generated by $k$ rows of matrix $G_m = M^{(m)}K^{\otimes m}$, where $\otimes m$ is $m$-fold Kronecker product of matrix with itself, and  $M^{(m)}$ is a digit-reversal permutation matrix, corresponding to mapping  
 $ \sum_{i = 0}^{m-1}j_il^i \rightarrow  \sum_{i = 0}^{m-1}j_{m-1-i}l^i$, $j_i \in [l]$.
The encoding  scheme is given by
$ c_0^{n-1}= u_0^{n-1}G_m$,
where $u_i,i\in \mathcal F$, are  set to some pre-defined values, e.g. zero (frozen symbols),  $|\mF| = n - k$, and the remaining values $u_i$ are set to the payload data. 

It is possible to show that a binary input memoryless channel $W$ together with matrix $G_m$ gives rise to bit subchannels $W_{m,K}^{(i)}
(y_0^{n-1},u_0^{i-1}|u_i)$ with capacities approaching $0$ or $1$. The fraction of noiseless subchannels approaches $I(W)$ \cite{korada2010polar}. Selecting  $\mF$ as the set of indices of low-capacity subchannels enables almost error-free communication. 

It is convenient to define probabilities 
\begin{align*}
 W^{(i)}_{m,K}(u_0^{i}|y_0^{n-1})=
 \frac{W_{m,K}^{(i)}(y_0^{n-1},u_0^{i-1}|u_i)}{2W(y_0^{n-1})}= 
 \quad \quad \quad \quad \quad \quad\nonumber\\
 \sum_{u_{i+1}^{n-1}}W_{m,K}^{(n-1)}(u_0^{n-1}|y_0^{n-1})=
 \sum_{u_{i+1}^{n-1}}\prod_{i = 0}^{n-1}W((u_0^{n-1}G_m)_i|y_i).
 \end{align*}
We further define  $\W^{(i)}_{m}(u_0^{j}|y_0^{n-1}) = W^{(i)}_{m,K}(u_0^{i}| y_0^{n-1})$, where $K$ should be clear from the context. 

Due to the recursive structure of $G_m$,
one has
\begin{align}
\label{mKernProb}
\W^{(sl+\phi)}_{m}(u_0^{sl+\phi}|y_0^{n-1}) = 
  \quad \quad \quad \quad \quad \quad \quad \quad \quad \quad \quad \nonumber\\ 
\sum_{u_{sl+\phi+1}^{l(s+1)-1}} \prod_{j = 0}^{l-1} \W_{m-1}^{(s)}
(\theta_K[u_0^{l(s+1)-1},j]|y_{j\frac{n}{l}}^{(j+1)\frac{n}{l}-1}),
\end{align}
where $\theta_K[u_0^{(s+1)l-1},j]_r = (u_{lr}^{l(r+1)-1}K)_j, r \in [s+1], \phi \in [l]$.

 At the receiver side, the successive cancellation (SC) decoding algorithm makes estimates 
 \begin{equation*}
 \label{mSCProb}
 \widehat u_i=\begin{cases}\arg\max_{u_i\in \F_2} \W_m^{(i)}(\widehat u_0^{i-1}.u_i|y_0^{n-1}), &i\notin\mF,\\
\text{the frozen value of $u_i$}&i\in \mF.
\end{cases}
\end{equation*}
 
\subsection{Kernel processing}
 \label{s:kernelproc}
Decoding algorithms for polar codes require one to compute the probabilities $\W^{(i)}_{m}(u_0^{i}|y_0^{n-1})$ for a given polarization transform $G_m$. Since these probabilities are computed recursively \eqref{mKernProb}, we assume for simplicity that $m=1$.
The corresponding task is referred to as {\em kernel processing}\footnote{Sometimes it is also referred to as kernel marginalization \cite{bioglio2018marginalization}.}.
The probabilities for one layer of the polarization transform are given by
\begin{equation}
\W_1^{(\phi)}(u_0^\phi|y_0^{l-1}) = 
\sum_{u_{\phi+1}^{l-1} \in \F_2^{l-\phi-1}} \W^{(l-1)}_{1}(u_0^{l-1}|y_0^{l-1}).
\label{mKernelWExact}
\end{equation}
The value of $\phi$ is referred to as processing \textit{phase}, while vector $u_0^\phi$ is referred to as a \textit{path}. Computing these probabilities reduces to soft-output decoding of nonsystematically encoded codes generated by the last $l-\phi-1$ rows of $K$.  
This problem was considered in 
\cite{griesser2002aposteriori}.

We introduce approximate  probabilities 
\begin{equation}
\widetilde \W_1^{(\phi)}(u_0^\phi|\mathbf y_0^{l-1}) = \max_{u_{\phi+1}^{l-1} \in \F_2^{l-\phi-1}} \W^{(l-1)}_{1}(u_0^{l-1}|y_0^{l-1})
\label{mKernelWApprox}
\end{equation}
They were shown to provide a substantial reduction of the complexity of polar codes sequential decoding \cite{miloslavskaya2014sequentialBCH,miloslavskaya2014sequential}. 

Decoding can be implemented using the LLRs  of the approximate probabilities \eqref{mKernelWApprox}
\begin{equation}
\bS_{1, \phi}   = \ln \frac{\widetilde \W_1^{(\phi)}(u_0^{\phi-1}. 0|y_0^{l-1})}{\widetilde \W_1^{(\phi)}(u_0^{\phi-1}. 1|y_0^{l-1})} =R(0)-R(1), 
\label{mGenLLR}
\end{equation}
 where $R(a) = \max_{u_{\phi+1}^{l-1}}\ln \W^{(l-1)}_{1}(u_0^{\phi-1}.a.u_{\phi+1}^{l-1}|y_0^{l-1})$. 

The above expression means that $\bS_{1, \phi} $ can be computed by performing the maximum likelihood (ML) decoding of the coset of a code generated by last $l-\phi-1$ rows of the kernel $K$, assuming that all $u_j,\phi<j<l,$ are equiprobable.
The particular coset representative is given by the product of  $u_0^{\phi-1}$ and matrix consisting of top $\phi$ rows of kernel $K$.

\subsection{Processing of Arikan matrix}
\label{sLLRSimple}
Straightforward evaluation of \eqref{mGenLLR} for an arbitrary kernel has complexity $O(l2^l)$. However, there is a simple 
recursive procedure for computing \eqref{mGenLLR} for Arikan matrix 
$
F_t = \left(
\arraycolsep=1.15pt\def\arraystretch{0.5}
\begin{array}{cc}
1&0\\
1&1
\end{array} \right)^{\otimes t}.
$

Let $l = 2^t$. Consider encoding scheme 
\begin{equation*} 
c_0^{l-1} = v_0^{l-1}F_t.
\end{equation*}
Similarly to \eqref{mKernelWApprox}, define approximate probabilities $$\widetilde W_t^{(i)}(v_0^i|y_0^{l-1})=
\max_{v_{i+1}^{l-1} \in \F_2^{l-i-1}} W_t^{(l-1)}(v_0^{l-1}|y_0^{l-1})
$$
and modified log-likelihood ratios $$S_t^{(i)}(v_0^{i-1},y_0^{l-1})=\log\frac{\widetilde W_t^{(i)}(v_0^{i-1}.0|y_0^{l-1})}{\widetilde W_t^{(i)}(v_0^{i-1}.1|y_0^{l-1})}.$$

It can be seen \cite{trifonov2018score} that 
\begin{align}
S_{\lambda}^{(2i)}(v_0^{2i-1},y_0^{N-1})
=&Q(a,b)=\sgn (a)\sgn (b)\min(|a|,|b|)\label{mMinSum1}\\
S_{\lambda}^{(2i+1)}(v_0^{2i},y_0^{N-1})=&P(a,b,v_{2i})=(-1)^{v_{2i}}a+b
,\label{mMinSum2}\\
a=&S_{\lambda-1}^{(i)}(v_{0,e}^{2i-1}\oplus v_{0,o}^{2i-1},y_{0,e}^{{N}-1}), \label{mRecA}\\
b=&S_{\lambda-1}^{(i)}(v_{0,o}^{2i-1},y_{0,o}^{N-1}),\label{mRecB}
\end{align}
where $N=2^{\lambda}$. The initial values 
for this recursion 
are given by $S_0^{(0)}(y_i)=\log\frac{W(0|y_i)}{W(1|y_i)}$. 
These expressions can be 
immediately 
 recognized as the min-sum approximation of the SCL decoding \cite{balatsoukasstimming2015llrbased}. However, these
are also the exact values, which reflect the probability of the most likely valid continuation of a given path $v_0^{i-1}$. 

Observe that one does not need to compute all values given by the recursion \eqref{mMinSum1}--\eqref{mMinSum2} at each phase. It is possible to reuse intermediate LLRs $S_\lambda^{(i)} = S_\lambda^{(i)}(v_0^{i-1}|y_0^{N-1})$, obtained in previous phases. At phase $i > 0$ one needs to compute only values $S_{t - j}^{(\lfloor i/2^{s - j}\rfloor)}$, $0 \leq j \leq s$, where $s$ is the largest integer such that $2^s$ divides $i$. Thus, the complexity of computing $S_t^{(i)}$ is $\sum_{j=0}^s 2^{j}$ operations, $i > 0$. See \cite{miloslavskaya2014sequential} or \cite{trofimiuk2020fast} for details. 

The log-likelihood of a path $v_{0}^i$ can be obtained \cite{trifonov2018score} as 
\begin{align}
R_y(v_0^i) = R(v_0^i|y_0^{l-1})=
\log\widetilde W_t^{(i)}(v_0^i|y_0^{l-1})
\quad \quad \quad \quad \nonumber\\
=R_y(v_0^{i-1})+\tau\left(S_t^{(i)}(v_0^{i-1},y_0^{l-1}),v_i\right),
\label{f:PathScore}
\end{align}
where $R_y(\epsilon)$ can be set to $0$, $\epsilon$  is an empty sequence, and $\tau(S,v)=\begin{cases}
0,&\sgn(S)=(-1)^v\\
-|S|,&\text{otherwise}.
\end{cases}$ 

\subsection{Fundamental parameters of polar codes}
\subsubsection{Rate of polarization}
%
%

The rate of polarization $E(K)$ shows how fast bit subchannels of $K^{\otimes m}$ approach either almost noiseless or pure noise channel with $n = l^m$. Suppose we construct $(n,k)$ polar code $\mathcal C$ with kernel $K$. Let $P_e(n)$ be a block error probability of SC decoding of  $\mathcal C$ under transmission over $W$. It was shown \cite{korada2010polar}, that if $k/n <\ I(W)$ and $\beta < E(K)$, then for sufficiently large $n$
the probability $P_e(n)$ can be bounded as 
$
P_e(n) \leq 2^{-n^\beta}.
$ 

The rate of polarization is independent of channel $W$. The method of its computation is proposed in \cite{korada2010polar}.
 It is possible to obtain $l\times l$ kernels with the rate of polarization arbitrarily close to 1  by increasing the kernel dimension $l$ to infinity \cite{korada2010polar}. Explicit constructions of kernels with high rate of polarization are provided in \cite{presman2015binary,lin2015linear}.
\subsubsection{Scaling exponent}
Another crucial property of polarization kernels is the \textit{scaling exponent}.
Let us fix a binary discrete memoryless channel $W$ of capacity $I(W)$ and a desired block error probability $P_e$. Given $W$ and $P_e$,
suppose we wish to communicate at rate $I(W) - \Delta$ using a family of $(n,k)$ polar codes with kernel $K$. The value of $n$
scales as $O(\Delta^{- \mu(K)})$, where the constant $\mu(K)$ is known as the scaling exponent \cite{fazeli2014scaling}. The scaling exponent depends on the channel. Unfortunately, the algorithm of its computation is only known for the case of binary erasure channel \cite{hassani2014finitelength}, \cite{fazeli2014scaling}. It is possible to show \cite{pfister2016nearoptimal,fazeli2020binary} that there exist $l\times l $ kernels $K_l$, such that $\lim_{l\rightarrow \infty}\mu(K_l)=2,$
i.e. the corresponding polar codes provide optimal scaling behaviour. 
Constructions of the kernels with good scaling exponent are provided in 
\cite{trofimiuk2019construction32,yao2019explicit}.

\section{Window Processing}
\label{sConventionalWindow}

In this section we present a detailed description of the \textit{window processing} (WP) algorithm for large polarization kernels. 
The basic idea of WP is to calculate the input symbols probabilities of the desired kernel $K$ via probabilities of Arikan matrix $F_t$, which are easy to compute. This idea was originally proposed in  \cite{trifonov2014binary} and reinvented in \cite{buzaglo2017permuted}.

We also present an implementation of WP in LLR domain of 
approximated probabilities \eqref{mGenLLR}.

\subsection{Relation between the input vectors of polarizing transforms}
\label{ssConventionalWindow}
WP algorithm calculates the probabilities $\W^{(\phi)}_1(u_0^{\phi}|y_0^{l-1} )$ of input symbols $u_0^{\phi}$  for $l \times l, l = 2^t$, kernel $K$ using probabilities $W^{(i)}_{t}(v_0^{i}|y_0^{l-1}), i \geq \phi$ for Arikan matrix $F_t$. This operation requires establishing the relation between the input vectors $u$ and $v$ of polarizing transforms $K$ and $F_t$. 

Indeed, since $K$ and $F_t$ are invertible, we can write $TK = F_{t}$, where $T$ is referred to as the \textit{transition matrix}. Hence, any vector can be represented as 
$c_{0}^{l-1} = v_0^{l-1} F_{t} =u_{0}^{l-1}K$. This implies that $u_0^{l-1} = v_0^{l-1}T,$
or
\begin{equation}
\label{mInputTransitionExp}
u_\phi = \sum_{s = 0}^{l-1} v_s T[s,\phi] = \sum_{\substack{s = 0 \\T_{[s,\phi]} = 1}}^{\tau_\phi} v_s,
\end{equation}
where $\tau_\phi$ denotes the position of the last non-zero symbol in the $\phi$-th column of $T$. We rewrite \eqref{mInputTransitionExp} to obtain
\begin{equation}
\label{mFirstV}
v_{\tau_\phi} = u_\phi + \sum_{\substack{s= 0 \\T_{[s,\phi]} = 1}}^{\tau_\phi-1} v_s
\end{equation}

However, some $\tau_\phi$ might be equal. This means that some $v_i, i\in[l]$, are not explicitly defined by \eqref{mFirstV}. Nevertheless, it is possible to use linear operations to transform \eqref{mFirstV} to derive the equations for all $v_i, i \in [l]$. 

Indeed, the vectors $u_0^{l-1}$ and $v_0^{l-1}$ satisfy the  equation
$$\Theta'(u_{l-1}, \dots , u_1, u_0, v_0, v_1 , \dots , v_{l-1})^{T}=0,$$
where $\Theta'=(\mathbb T\quad I)$, and $l\times l$ matrix $\mathbb T$ is obtained by transposing $T^{-1}$ and reversing the order of columns in the obtained matrix. By applying elementary row operations, the matrix $\Theta'$
can be transformed into a minimum-span form $\Theta$, such that the  first and last non-zero elements of the $\phi$-th row are located in  columns $\phi$ and  $z_\phi$, respectively, where all $z_\phi$ are distinct. This enables one to obtain the symbols of 
the vector 
 $u$ as 
\begin{align}
\label{fTransformMinSpan}
u_{\phi}=\sum_{s=0}^{\phi-1} u_s\Theta_{l-1-\phi,l-1-s} +
\sum_{j=0}^{\omega_\phi}v_j \Theta_{l-1-\phi, l+j},
\end{align}
where $\omega_\phi=z_{l-1-\phi}-l$. Using \eqref{fTransformMinSpan}, we write the equations for the vector $v_0^{l-1}$:
\begin{equation}
\label{fTransformMinSpanV}
v_{\omega_\phi} = \sum_{s=0}^{\phi} u_s\Theta_{l-1-\phi,l-1-s} +
\sum_{j=0}^{\omega_\phi-1}v_j \Theta_{l-1-\phi, l+j}.
\end{equation}

In this paper we consider kernels with all $\tau_\phi, \phi \in [l]$, distinct. In this case, we do not need to compute the matrix $\Theta$ to obtain \eqref{fTransformMinSpan}. For such kernels we have $\omega_\phi = \tau_\phi,\phi\in[l]$, and can immediately use \eqref{mInputTransitionExp} to evaluate $u$ from $v$.

\begin{figure}[t]
\hspace{-0.2cm}
\scalebox{0.85}{
\parbox{0.2\textwidth}
{ 
$
\arraycolsep=1.35pt\def\arraystretch{0.7}
\begin{array}{cc}
{
 \begin{array}{c}
K_{16}', E = 0.51828, \mu = 3.346\\
\left({\begin{array}{cccccccccccccccc}
1&0&0&0&0&0&0&0&0&0&0&0&0&0&0&0\\
1&1&0&0&0&0&0&0&0&0&0&0&0&0&0&0\\
1&0&1&0&0&0&0&0&0&0&0&0&0&0&0&0\\
1&0&0&0&1&0&0&0&0&0&0&0&0&0&0&0\\
1&0&0&0&0&0&0&0&1&0&0&0&0&0&0&0\\
1&1&0&0&0&0&0&0&1&1&0&0&0&0&0&0\\
1&0&1&0&0&0&0&0&1&0&1&0&0&0&0&0\\
1&1&1&1&0&0&0&0&0&0&0&0&0&0&0&0\\
1&0&0&0&1&0&0&0&1&0&0&0&1&0&0&0\\
0&1&1&0&1&1&0&0&1&0&1&0&0&0&0&0\\
1&1&0&0&1&0&1&0&0&1&1&0&0&0&0&0\\
1&1&1&1&1&1&1&1&0&0&0&0&0&0&0&0\\
1&1&1&1&0&0&0&0&1&1&1&1&0&0&0&0\\
1&1&0&0&1&1&0&0&1&1&0&0&1&1&0&0\\
1&0&1&0&1&0&1&0&1&0&1&0&1&0&1&0\\
1&1&1&1&1&1&1&1&1&1&1&1&1&1&1&1\\
\end{array}}\right)\\
 \end{array}
}&
{
\begin{array}{c}
K_{16}, E = 0.51828, \mu = 3.45\\
{\left({\begin{array}{cccccccccccccccc}
1&0&0&0&0&0&0&0&0&0&0&0&0&0&0&0\\
1&1&0&0&0&0&0&0&0&0&0&0&0&0&0&0\\
1&0&1&0&0&0&0&0&0&0&0&0&0&0&0&0\\
1&1&1&1&0&0&0&0&0&0&0&0&0&0&0&0\\
1&0&0&0&1&0&0&0&0&0&0&0&0&0&0&0\\
1&0&0&0&0&0&0&0&1&0&0&0&0&0&0&0\\
1&1&0&0&0&0&0&0&1&1&0&0&0&0&0&0\\
1&0&1&0&0&0&0&0&1&0&1&0&0&0&0&0\\
0&1&1&0&1&1&0&0&1&0&1&0&0&0&0&0\\
1&1&0&0&1&0&1&0&0&1&1&0&0&0&0&0\\
1&1&1&1&1&1&1&1&0&0&0&0&0&0&0&0\\
1&1&1&1&0&0&0&0&1&1&1&1&0&0&0&0\\
1&0&0&0&1&0&0&0&1&0&0&0&1&0&0&0\\
1&1&0&0&1&1&0&0&1&1&0&0&1&1&0&0\\
1&0&1&0&1&0&1&0&1&0&1&0&1&0&1&0\\
1&1&1&1&1&1&1&1&1&1&1&1&1&1&1&1\\
\end{array}}\right)}
\end{array}
}
\end{array}
$
}}
\caption{Large polarization kernels with improved polarization}
\label{f:Kernels16}
\end{figure}

Fig. \ref{f:Kernels16} presents $16 \times 16$  kernels $K_{16}'$ and $K_{16}$, which were constructed by algorithm given in [26]. They have improved polarization properties and admit low complexity WP. The expressions \eqref{mInputTransitionExp} for $K_{16}'$ and $K_{16}$ are given in Table \ref{tDecWin16}.

\begin{table}[ht]
\caption{Transition matrices of $K_{16}', K_{16}$ kernels}
\label{tDecWin16}
\centering
\footnotesize
\scalebox{0.95}{
\begin{tabular}{|l|p{0.1\textwidth}|c|c||p{0.1\textwidth}|c|c|}
\hline
\multirow{2}{*}{$\phi$}&\multicolumn{3}{c||}{$K_{16}'$}&\multicolumn{3}{c|}{$K_{16}$}\\\cline{2-7}
   & $u_\phi $         & $\mathcal D_\phi$ & Cost& $u_\phi$      & $\mathcal D_\phi$ & Cost\\ \hline
0  & $v_0$              & $\set{}$ &15& $v_0$                      &$\set{}$&15\\ \hline
1  & $v_1$              & $\set{}$ &1& $v_1$                      &$\set{}$  &1\\ \hline
2  & $v_2$              & $\set{}$ &3& $v_2$                      & $\set{}$ &3\\ \hline
3  & $v_4$              & $\set{3}$ &21& $v_3$                     &$\set{}$  &1\\ \hline
4  & $v_8$              & $\set{3,5,6,7}$ &127& $v_4$               &$\set{}$  &7\\ \hline
5  & $v_6\oplus v_9$      & $\set{3,5,6,7}$ &48& $v_8$               & $\set{5,6,7}$ &67\\ \hline
6  & $v_5\oplus v_6 \oplus v_{10} $   & $\set{3,5,6,7}$ &95& $v_6\oplus v_9$       & $\set{5,6,7}$ &24\\ \hline
7  & $v_3$              & $\set{5,6,7}$ &1& $v_5\oplus v_6\oplus v_{10}$      & $\set{5,6,7}$ &47\\ \hline
8  & $v_{12}$           & $\set{5,6,7,11}$ &127& $v_{5}$              & $\set{6,7}$ &1\\ \hline
9  & $v_{5}$           & $\set{6,7,11}$ &1& $v_{6}$             & $\set{7}$ &1\\ \hline
10 & $v_{6}$              & $\set{7,11}$ &1& $v_7$                  & $\set{}$ &1\\ \hline
11 & $v_7$              & $\set{11}$ &1& $v_{11}$                 & $\set{}$ &1\\ \hline
12 & $v_{11}$           & $\set{}$ &1& $v_{12}$                   & $\set{}$ &7\\ \hline
13 & $v_{13}$           & $\set{}$ &1& $v_{13}$                   & $\set{}$ &1\\ \hline
14 & $v_{14}$           & $\set{}$ &3& $v_{14}$                   & $\set{}$ &3\\ \hline
15 & $v_{15}$           & $\set{}$ &1& $v_{15}$                   & $\set{}$ &1\\ \hline
\end{tabular}
}
\end{table}

\subsection{Decoding window}
In the previous section we discussed the relation between input vectors $u$ and $v$ of polarizing transforms $K$ and $F_t$. In this section we use this relation to compute the probabilities of $K$ via the probabilities of $F_t$.

Using \eqref{fTransformMinSpan} we can reconstruct $u_0^\phi$ from $v_0^{h_\phi}$, where  $h_\phi = \underset{0 \leq \phi' \leq \phi}{\max} \tau_{\phi'}$. Using this fact, we want to express the probability $\W^{(\phi)}_1(u_0^{\phi}|y_0^{l-1})$ via $W^{(h_\phi)}_{t}(v_0^{h_\phi}|y_0^{l-1})$. However, if $h_\phi > \phi$, then some values of $v_0^{h_\phi}$ are independent from $u_0^{\phi}$ and, therefore, unknown. This suggests that to compute $\W^{(\phi)}_1(u_0^{\phi}|y_0^{l-1})$ we need to consider the probabilities $W^{(h_\phi)}_{t}(v_0^{h_\phi}|y_0^{l-1})$ for vectors $v_0^{h_\phi}$ with all possible values of unknown bits.
 
 Let
$$
\mathcal D_\phi = [h_\phi+1]\setminus  \{\omega_0,\omega_1,\dots,\omega_\phi\}
$$
 be a \textit{decoding window}, i.e. the set of indices of independent (from  $ u_0^{\phi}$) components of $v_0^{h_\phi}$. Note that 
 $$|\mathcal D_\phi| = |[h_\phi+1]| - |\{\omega_0,\omega_1,\dots,\omega_\phi\}| = h_\phi - \phi,$$ since all $\omega_\phi$ are distinct and $\{\omega_0,\omega_1,\dots,\omega_\phi\} \subseteq [h_\phi+1]$.

\begin{theorem}
\label{theoremWindow}
Let $K$ be an $l \times l, l = 2^t,$ polarization kernel. Then, the probability of the input vector $u_0^{\phi}$ of 
$K$ is given by\footnote{The method proposed in \cite{buzaglo2017permuted} is a special case of this approach. }
\begin{equation}
\label{mKernW}
\W^{(\phi)}_1(u_0^{\phi}|y_0^{l-1}) = \sum_{v_0^{h_{\phi}}\in \mZ_\phi^{(u_\phi)}}
W^{(h_\phi)}_{t}(v_0^{h_{\phi}}|y_0^{l-1}),
\end{equation}
where $\mZ_\phi^{(b)}$ is the set of vectors $v_0^{h_\phi}$, such as 
$v_s \in \bF_2, s\in \mD_\phi$, the values of $v_t, t \in [h_\phi+1] \backslash \mathcal D_\phi,$ are obtained according to expression \eqref{fTransformMinSpanV} under condition that $u_\phi = b$.
\end{theorem}
\begin{proof}
By definition \eqref{mKernelWExact}, we have
\begin{align}
&\W^{(\phi)}_1(u_0^{\phi}|y_0^{l-1}) = \sum_{u_{\phi+1}^{l-1}\in \F_2^{l-\phi-1}} 
\W^{(l-1)}_{1}(u_0^{l-1}|y_0^{l-1})=
\nonumber \\
&\sum_{u_{\phi+1}^{l-1} \in \F_2^{l-\phi-1}}\prod_{i = 0}^{l-1}W((u_0^{l-1}K)_i|y_{i})
\overset{a}{=} \nonumber \\ 
&\sum_{v_0^{h_{\phi}}\in\mZ_\phi^{(u_\phi)}, v_{h_\phi+1}^{l-1}\in \F_2^{l-h_\phi-1}}\prod_{i = 0}^{l-1}W(((v_0^{l-1}T)K)_i|y_{i})
\overset{b}{=}
\nonumber \\
&\sum_{v_0^{h_{\phi}}\in \mZ_\phi^{(u_\phi)}}\sum_{v_{h_\phi+1}^{l-1} \in \F_2^{l-h_\phi-1}} 
\prod_{i = 0}^{l-1}W((v_0^{l-1}F_t)_i|y_{i}) \overset{c}{=}
\nonumber \\
& \sum_{v_0^{h_{\phi}}\in\mZ_\phi^{(u_\phi)}}\sum_{v_{h_\phi+1}^{l-1} \in \F_2^{l-h_\phi-1}} 
 W_t^{(l-1)}(v_0^{l-1}|y_0^{l-1})=
 \nonumber\\
& \sum_{v_0^{h_{\phi}}\in\mZ_\phi^{(u_\phi)}} W^{(h_\phi)}_{t}(v_0^{h_{\phi}}|y_0^{l-1})\nonumber.
\label{mKernelLook}
\end{align}
Equality (a) follows from \eqref{fTransformMinSpan} and definition of $\mathcal Z_\phi^{(u_\phi)}$. Note that $|\mZ_\phi^{(u_\phi)}| \cdot |\F_2^{l-h_\phi-1}| = |\F_2^{l-\phi-1}|$.
Equality (b) follows from the definition $TK = F_t$. 
Equality (c) is due to
\eqref{mKernelWExact}.
\end{proof}

The expression \eqref{mKernW} is the WP algorithm in the probabilistic domain.  It  means that computation of the probability $\W^{(\phi)}_1(u_0^{\phi}|y_0^{l-1})$ requires considering $2^{|\mD_\phi|+1}$ paths $v_0^{h_\phi} \in \mZ_\phi^{(u_\phi)}$ in context of Arikan SC decoding and computing $2^{|\mD_\phi|+1}$ probabilities $W^{(h_\phi)}_{t}(v_0^{h_{\phi}}|y_0^{l-1})$ for each path. 

Let $\mathcal M(K)$ denote the $\max_{i \in [l]}|\mathcal D_i|$ for kernel $K$.  In general, one has $\mathcal M(K) = O(l)$ for an arbitrary  $K$. In this case, the complexity of  computing \eqref{mKernW} is given by $O(2^l)$ arithmetic operations. Fortunately, there are some methods \cite{trofimiuk2019construction32}, \cite{abbasi2020large} to construct the polarization kernels with the reduced $\mathcal M(K)$, which allow low complexity WP. 


\begin{example}
Consider computing $\W_1^{(6)}(u_0^{6}|y_0^{15})$ for $K_{16}'$. By definition \eqref{mKernelWExact} one should compute 
\begin{equation}
\label{fK16Example}
\W_1^{(6)}(u_0^{6}|y_0^{15}) = \sum_{u_{7}^{15} \in \F_2^9} \W^{(15)}_{1}(u_0^{15}|y_0^{15}).
\end{equation}

The value of $h_6 = 10$ indicates the largest index of the dependent (from $u_0^6$) element of $v$. It implies that we should consider the probabilities $W^{(10)}_{4}(v_0^{10}|y_0^{15})$  of paths $v_0^{10}$. However, in $v_0^{10}$ we have 4 independent (unknown) elements: $v_3$ and $v_5^7$. Indices of these elements form the decoding window $\mD_6 = \set{3,5,6,7}$.
The expressions for each $v_i$ are given by \eqref{mFirstV}. This enables us to construct the set $\mZ_6^{(u_6)}$ of paths $v_0^{10}$. 

In this example we have $\mZ_6^{(u_6)} = \{ v_0^{10}| v_i = \bF_2,i \in \mD_6, v_j = u_j, j \in [3], v_4 = u_3, v_8 = u_4, v_9 = u_6 \oplus v_5, v_{10} = v_5 \oplus v_6 \oplus u_6\}$ and $|\mZ_6| = 16$. Thus, according to Theorem \ref{theoremWindow},
$$\W^{(6)}_1(u_0^{6}|y_0^{15}) = \sum_{v_0^{10}\in\mZ_6^{(u_6)}}
W^{(10)}_{4}(v_0^{10}|y_0^{15}).$$
\end{example}

The decoding windows of $K_{16}', K_{16}$  are presented in Table \ref{tDecWin16} together with WP complexity of each phase. The detailed description of $K_{16}$ processing is presented in Section \ref{ssDescr16_2}. The computation of WP\ complexity of $K_{16}'$  is given in \cite{trofimiuk2018efficientArXiv}.
\subsection{Log-likelihood domain}
\label{ssLLRWind}
Similarly to \eqref{mKernelWApprox}, we can rewrite \eqref{mKernW} for the case of approximate probabilities
\begin{align}
\widetilde \W_1^{(\phi)}(u_0^\phi|y_0^{l-1}) = \max_{v_0^{h_{\phi}}\in 
\mZ_\phi^{(u_\phi)}}  \widetilde W^{(h_\phi)}_{t}(v_0^{h_{\phi}}|y_0^{l-1}) =  \nonumber \\ 
\max_{v_0^{h_{\phi}}\in\mZ_\phi^{(u_\phi)}} \max_{v_{h_\phi+1}^{l-1}} W^{(l-1)}_{t}(v_0^{l-1}|y_0^{l-1}).
\end{align} 
Using  \eqref{f:PathScore}, we can obtain the LLRs
\begin{align}
\bS_{1, \phi} = \max_{v_0^{h_{\phi}}\in\mathcal Z_{\phi}^{(0)}} R_y(v_0^{h_\phi}) -
\max_{v_0^{h_{\phi}}\in\mathcal Z_{\phi}^{(1)}} R_y(v_0^{h_\phi}).
\label{mKernLLR}
\end{align}

Calculation of LLRs $\bS_{1,\phi}$ via \eqref{mKernLLR} is referred to as the WP algorithm in LLR domain. Let 
$\mZ_\phi =  \mZ_\phi^{(0)} \cup \mZ_\phi^{(1)}$. The number of path scores to be computed  in \eqref{mKernLLR} is equal to $|\mZ_{\phi}| = 2^{|\mathcal D_\phi|+1}$. It is also possible to reuse the intermediate LLRs for each path $v_0^{h_{\phi}}$ (see Section \ref{sLLRSimple}).

Although the number of path scores $R_y(v_0^{h_\phi})$, $v_0^{h_\phi}$ $\in$ $\mZ_\phi$, is $2^{|\mathcal D_\phi|+1}$,  they can be computed with $2^{|\mathcal D_\phi|}$ operations. Namely, 
during the calculation of each $R_y(v_0^{h_\phi})$, both terms 
$\tau(S_{t}^{(h_\phi)}(v_0^{h_\phi-1},y_0^{l-1}),b)$, $b \in \bF_2$, arise. Due to the definition of $\tau$, there is such value of $a$ $\in$ $\bF_2$, that  
$\sgn (S_{t}^{(h_\phi)}(v_0^{h_\phi-1},y_0^{l-1})) = (-1)^{a}$, thus, $\tau(S_{t}^{(h_\phi)}(v_0^{h_\phi-1},y_0^{l-1}), a)$ $=$ $0$. It remains to compute only $\tau(S_{t}^{(h_\phi)}(v_0^{h_\phi-1},y_0^{l-1}),a \oplus 1)$. The number of such nonzero terms is $2^{|\mD_\phi|}$.

The equations \eqref{mKernLLR} and \eqref{mMinSum1}--\eqref{f:PathScore} can be immediately used to perform kernel processing. However, in the next sections we show how to substantially reduce the complexity of WP.

\section{Common Subexpressions Identification}
\label{s:CSE}
In this section we propose a simplified version of WP algorithm. Conventional WP considers several paths in the context of Arikan SC\ decoding. For each path, the algorithm recursively computes its LLR (which is used to obtain the path score). It turns out that the intermediate LLRs induced by this recursion can be equal for different paths. Thus, one can compute only the unique intermediate LLRs, and, therefore, substantially reduce the processing complexity.
\subsection{Motivation}
\label{s:Motivation}
Suppose we want to compute 
$\phi$-th input symbol LLR $\bS_{1,\phi}$ of the $l \times l,l=2^t,$ polarization kernel $K$ by WP method.
It computes several scores $R_y(v_0^{h_\phi})$ of paths $v_0^{h_\phi}$, considered in the context of Arikan SC decoding. The number of such paths is determined by the size of the $\phi$-th decoding window $\mathcal D_\phi$. The values $\phi$ and $h_\phi$ are referred to as \textit{external phase} and \textit{internal phase}. 

According to \eqref{f:PathScore}, to obtain a single path score $R_y(v_0^{h_\phi})$ one needs to compute the LLR $S_t^{(h_\phi)} = S_t^{(h_\phi)}(v_0^{h_\phi-1},y_0^{l-1})$ and the path score $R_y(v_0^{h_\phi-1})$ from the previous phase $h_\phi-1$. The value $S_t^{(h_\phi)}$ is computed recursively via  \eqref{mMinSum1}--\eqref{mRecB}. On each layer $\lambda \in [t+1]$ of this recursion $M = 2^{t-\lambda}$
intermediate LLRs $S_{\lambda}^{(j)} = S_{\lambda}^{(j)}(\bar v_0^{j-1},\bar y_0^{N-1}), N = 2^{\lambda}$, are arising. The value $\bar v_0^{j-1}$ denotes a partial sum of some elements from $v_0^{h_\phi-1}$ and $\bar y_0^{N-1}$ is a particular subvector  of $y_0^{l-1}$. 

It is possible to write $\bar v_0^{j-1}$ and $\bar y_0^{N-1}$ explicitly for all $S_\lambda^{(j)}$. 
We introduce an array $C_\lambda^{(j)}[\beta], \beta \in [M],$ of partial sums $\bar v_0^{j-1}$ at layer $\lambda$  
\begin{equation}
\label{fExlpC}
C_\lambda^{(j)}[\beta] = 
\begin{cases}
 (c_\beta^{(0)},c_\beta^{(1)},\dots, c_\beta^{(j-1)}),&j > 0,\\
 \epsilon,&  \text{otherwise},
 
\end{cases}
\end{equation}
where $c^{(i)} =v_{M \cdot i}^{M \cdot i+M-1}F_{t-\lambda}, i \in [j]$.
We also define an array $S^{(j)}_\lambda[\beta], \beta \in [M]$, of the intermediate LLRs $S_\lambda^{(j)}(\bar v_0^{j-1},\bar y_0^{N-1})$ 
\begin{equation}
\label{fSExact}
S^{(j)}_\lambda[\beta] =
S_\lambda^{(j)}(C_\lambda^{(j)}[\beta],(y_{\beta},y_{\beta +M},\dots, y_{\beta+M(N-1)})).
\end{equation}

Similarly to \cite{miloslavskaya2014sequential}, these LLRs can be computed as
\begin{equation*}
S_\lambda^{(j)}[\beta]=
\begin{cases}
Q(S_{\lambda-1}^{(\hat j)}[\beta],S_{\lambda-1}^{
(\hat j)}[\beta+N])
, j \text{ even},\\
P(S_{\lambda-1}^{(\hat j)}[\beta],S_{\lambda-1}^{(\hat j)}[\beta+N], 
(C_{\lambda}^{(j)}[\beta])_{j-1})
, j \text{ odd},
\end{cases}
\end{equation*}
where $\hat j = \lfloor j/2 \rfloor$. 

WP method requires calculation  of $2^{|\mD_\phi|}$ LLRs $S_t^{(h_\phi)}$ given by $v_0^{h_\phi-1} \in \mZ_\phi$. For convenience, we define a set $$\mathbb S^{(h)}_\phi = \set{S_t^{(h)}(v_0^{h-1},y_0^{l-1})|v_0^{h_\phi-1} \in \mZ_\phi}, h \leq h_\phi.$$
In a straightforward implementation, all intermediate LLRs 
$S^{(j)}_\lambda[\beta]$ should be separately computed for each $S_t^{(h_\phi)} \in \mathbb S^{(h_\phi)}_\phi$. It turns out that $S^{(j)}_\lambda[\beta]$ induced by recursive calculation of $S_t^{(h_\phi)}$ for different $v_0^{h_\phi-1} \in \mZ_\phi$  may be the same. For example, the number of distinct vectors $\bar v_0^{j-1}$ and length-$N$ subvectors $\bar y_0^{N-1}$ of $y_0^{l-1}$ arising in \eqref{mMinSum1}--\eqref{mMinSum2}  is at most $2^j$ and $2^{t-\lambda}$, respectively. Thus, the number of distinct pairs $(\bar v_0^{j},\bar y_0^{N-1}) \leq 2^{j+t-\lambda}$, which can be even less than $2^{|\mD_\phi|}$.

We propose to compute $S^{(j)}_\lambda[\beta] = S_{\lambda}^{(j)}(\bar v_0^{j},\bar y_0^{N-1})$ only for distinct pairs ($\bar v_0^{j}, y_0^{N-1}$) arising at each layer of the recursion \eqref{mMinSum1}--\eqref{mRecB}. We denote by $X_\lambda = X_\lambda^{( h_\phi)}$, $\lambda \in [t+1]$, the set of these distinct pairs $(\bar v_0^{j},\bar y_0^{N-1})$ and refer it to as a set of \textit{common subexpressions} (CSE). 
In the case when $h_{\phi} - 1$ is greater than $h_{\phi - 1}$, one needs to compute several sets $\mathbb S^{(h)}_\phi$ using CSE\ for each  phase $h$, where  $h_{\phi-1}<h < h_{\phi}$.



\subsection{The CSE identification algorithm}
\label{ssCSEIdent}

The sets $X_\lambda$ of CSE can be identified offline for a given polarization kernel $K$ of size $l = 2^t$, external phase $\phi$ and  phase $h$, $h_{\phi - 1} < h < h_{\phi +\ 1}$.  To do that, one should run the procedure \textit{GetCSEPairs}$(h,\phi)$ illustrated in Alg. \ref{fCSEAlg}. This procedure iteratively enumerates the pairs $(\bar v_0^{j-1}, \bar y_0^{N-1})$ according to the recursion in \eqref{mMinSum1}--\eqref{mRecB}, and stores the unique ones only. It starts with pairs $(v_0^{h-1}|y_0^{l-1})$ in symbolic form, where $v_0^{h-1} \in \widehat \mZ_{\phi,h}$, 
$\widehat \mZ_{\phi,h} = \set{v_0^{h - 1}|v_i \in \set{0,1}, i \in \mathcal D_\phi, u_j = \widehat u_j, j \in [\phi]}$. Recall that $\widehat u_i$ is already estimated by SC\ decoding. Observe that $X_0 = \set{S_0^{(0)}(\epsilon, y_i),i\in[l]}$ for any $h$.
\begin{algorithm}[ht]
$\widehat \mZ_{\phi,h} = \set{v_0^{h - 1}|v_i \in \set{0,1}, i \in \mathcal D_\phi, u_j = \widehat u_j, j \in [\phi]}$\;\label{initZ}
$y_0^{l-1} \gets (y_0,y_1,\dots, y_{l-1})$ in symbolic form\;
$X_t \gets \set{(v_0^{h-1},y_0^{l-1})|v_0^{h - 1} \in \widehat \mZ_{\phi,h}}$\;\label{initX}
$I \gets h$\;
\For{$\lambda \gets t $ \textbf{downto} $1$}{
$X_{\lambda-1} \gets \emptyset; N \gets 2^\lambda$\;
$I' \gets I - (I \mod 2)$\;\label{r6}
\For{\textbf{each} $(\bar v_0^{I-1},\bar y_0^{N-1})$ \textbf{in} $X_\lambda$}{
$X_{\lambda-1} \gets
 X_{\lambda - 1} \cup \set{(v_{0,e}^{I'-1}\oplus v_{0,o}^{I'-1},y_{0,e}^{{N}-1})}$\;\label{r8}
$X_{\lambda-1} \gets X_{\lambda - 1} \cup \set{(v_{0,o}^{I'-1},y_{0,o}^{N-1})}$\;\label{r9}
}
$I \gets \lfloor I/2\rfloor$\;
}
\Return $\set{X_\lambda|\lambda \in [t+1]}$
\caption{GetCSEPairs($h,\phi$)}
\label{fCSEAlg}
\end{algorithm}


\subsection{LLR computation}
\label{ssCSELLR}

\begin{algorithm}[ht]
$S_0^{(0)}(\epsilon, y_i)\gets\log\frac{W(0|y_i)}{W(1| y_i)}, i \in [l]$\;

\For{$\lambda \gets 1$ \textbf{to} $t$}
{
$I \gets \lfloor h/2^{t-\lambda} \rfloor$ \; 
$I' \gets I - (I \mod 2)$\;
$N \gets 2^{t-\lambda}$\;
\For{\textbf{each} $(\bar v_0^{I-1},\bar y_0^{N-1})$ \textbf{in} $X_\lambda$}{
$a \gets S_{\lambda-1}^{(I'/2)}(v_{0,e}^{I'-1}\oplus v_{0,o}^{I'-1},y_{0,e}^{{N}-1})$\;
$b \gets S_{\lambda-1}^{(I'/2)}(v_{0,o}^{2i-1},y_{0,o}^{N-1})$\;
\If{$I$ \textbf{is even}}{
$S_\lambda^{(I)}(\bar v_0^{I-1},\bar y_0^{N-1}) \gets \sgn (a)\sgn (b)\min(|a|,|b|)$\;
}
\Else{
\text{evaluate $\bar v_{I}$ using }$\widehat u_0^{\phi-1}$ \text{as }$\bar c$\; \label{lcEval}
$S_\lambda^{(I)}(\bar v_0^{I-1},\bar y_0^{N-1}) \gets (-1)^{\bar c}a+b$\;}\label{lpEval}
}
}
\Return{$\mathbb S^{(h)}_\phi$}
\caption{ComputeLLRs($X,h,\phi$)}
\label{fComputeLLRsAlg}

\end{algorithm}
Finally, to obtain the LLRs $\mathbb S^{(h)}_\phi$ at the external phase $\phi$ of the kernel processing, one should run \textit{ComputeLLRs}$(X,h,\phi)$, presented in Alg. \ref{fComputeLLRsAlg}. It uses CSE pairs $(\bar v_0^{j-1}, \bar y_0^{N-1}) \in X_\lambda, \lambda \in \set{1,\dots,t},$ and iteratively computes all intermediate LLRs $S_\lambda^{(j)}(\bar v_0^{j-1}, \bar y_0^{N-1})$ corresponding to these pairs.

We would like to emphasise that CSE pairs $(\bar v_0^{j-1}, \bar y_0^{N-1}) \in X_\lambda$ are used in symbolic form, while the corresponding LLRs $S_\lambda^{(j)}(\bar v_0^{j-1}, \bar y_0^{N-1})$ are obtained as evaluated numbers in Alg. \ref{fComputeLLRsAlg}. The pairs $(\bar v_0^{j-1}, \bar y_0^{N-1})$ are used as indices of $S_\lambda^{(j)}$.  We assume that one can access any value $S_\lambda^{(j)}$ in constant time, since no sorting is required in Alg. \ref{fComputeLLRsAlg}. 
In a software implementation, one can replace the pairs $(\bar v_0^{j-1}, \bar y_0^{N-1})$ by integer indices and consider arrays of $S_\lambda^{(j)}$.

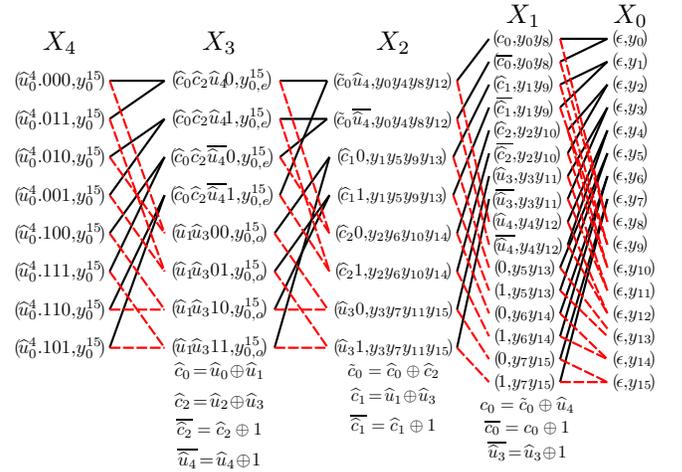
\begin{figure}
\centering
\scalebox{2.03}{
\begin{tikzpicture}
\begin{scope}[scale=0.5, shift={(-1.1,0)}]
\node[scale=0.5] at (0.05,0.75) {$X_4$};
\node[scale=0.35] at (0.05,0.25) {$(\!\widehat u_0^4. 000 , \!y_0^{15}\!)$};
\node[scale=0.35] at (0.05,-1.25) {$(\!\widehat u_0^4. 001 , \!y_0^{15}\!)$};
\node[scale=0.35] at (0.05,-0.75) {$(\!\widehat u_0^4. 010 , \!y_0^{15}\!)$};
\node[scale=0.35] at (0.05,-0.25) {$(\!\widehat u_0^4. 011 , \!y_0^{15}\!)$};
\node[scale=0.35] at (0.05,-1.75) {$(\!\widehat u_0^4. 100 , \!y_0^{15}\!)$};
\node[scale=0.35] at (0.05,-3.25) {$(\!\widehat u_0^4. 101 , \!y_0^{15}\!)$};
\node[scale=0.35] at (0.05,-2.75) {$(\!\widehat u_0^4. 110, \!y_0^{15}\!)$};
\node[scale=0.35] at (0.05,-2.25) {$(\!\widehat u_0^4. 111, \!y_0^{15}\!)$};
\end{scope}
\begin{scope}[shift={(-1.48,-2.15)},scale=0.5]
\node[scale=0.5] at (4,5.05) {$X_3$};
\node[scale=0.35] at (4.035,4.55) {$(\!\widehat c_0 \widehat c_2 \widehat u_4\!0,\! y_{0,e}^{15}\!)$};
\node[scale=0.35] at (4.035,4.05) {$(\!\widehat c_0 \widehat c_2 \widehat u_4\!1,\! y_{0,e}^{15}\!)$};
\node[scale=0.35] at (4,2.55) {$(\!\widehat u_1 \! \widehat u_3  00,\! y_{0,o}^{15}\!)$};
\node[scale=0.35] at (4,2.05) {$(\!\widehat u_1 \! \widehat u_3 01,\! y_{0,o}^{15}\!)$};
\node[scale=0.35] at (4.035,3.55) {$(\!\widehat c_0 \widehat c_2 \overline{\widehat u_4}0,\! y_{0,e}^{15}\!)$};
\node[scale=0.35] at (4.035,3.05) {$(\!\widehat c_0 \widehat c_2  \overline{\widehat u_4} 1,\! y_{0,e}^{15}\!)$};
\node[scale=0.35] at (4,1.55) {$(\!\widehat u_1 \! \widehat u_3 10,\! y_{0,o}^{15}\!)$};
\node[scale=0.35] at (4,1.05) {$(\!\widehat u_1 \! \widehat u_3 11,\! y_{0,o}^{15}\!)$};
\node[scale=0.35] at (4,0.75) {$\widehat c_0\!=\!\widehat u_0\!\oplus\!\widehat u_1$};
\node[scale=0.35] at (4,0.35) {$\widehat c_2\!=\!\widehat u_2\!\oplus\!\widehat u_3$};
\node[scale=0.33] at (4,0) {$\overline{\widehat c_2} = \widehat c_2 \oplus 1$};
\node[scale=0.35] at (4,-0.4) {$\overline{\widehat u_4}\!=\!\widehat u_4\!\oplus\!1$};
\end{scope}
\begin{scope}[shift={(-2.32,-2.6)},scale=0.5]
\node[scale=0.5] at (7.955,5.95) {$X_2$};
\node[scale=0.33] at (7.95,5.45) {$(\!\tilde  c_0\widehat u_4,\!y_0y_4y_{8}y_{12}\!)$};
\node[scale=0.33] at (7.95,4.95) {$(\!\tilde  c_0 \overline{\widehat u_4},\!y_0y_4y_{8}y_{12}\!)$};
\node[scale=0.33] at (7.96,3.45) {$(\!\widehat  c_20,\!y_2y_6y_{10}y_{14}\!)$};
\node[scale=0.33] at (7.96,2.95) {$(\!\widehat  c_21,\!y_2y_6y_{10}y_{14}\!)$};
\node[scale=0.33] at (7.94,4.45) {$(\!\widehat  c_1 0,\!y_1y_5y_{9}y_{13}\!)$};
\node[scale=0.33] at (7.94,3.95) {$(\!\widehat  c_1 1,\!y_1y_5y_{9}y_{13}\!)$};
\node[scale=0.33] at (7.96,2.45) {$(\!\widehat u_3 0,\!y_3y_7y_{11}y_{15}\!)$};
\node[scale=0.33] at (7.96,1.95) {$(\! \widehat u_31,\!y_3y_7y_{11}y_{15}\!)$};
\node[scale = 0.33] at (7.96,1.65) {$\tilde c_0 =\widehat c_0 \oplus \widehat c_2 $};
\node[scale = 0.33] at (7.96,1.32) {$\widehat  c_1\!=\!\widehat u_1\!\oplus\!\widehat u_3$};
\node[scale=0.33] at (7.96,0.95) {$\overline{\widehat c_1} = \widehat c_1 \oplus 1$};
\end{scope}
\begin{scope}[shift={(-1.49,-2.6)}]
\node[scale=0.5] at (4,3.15) {$X_1$};
\node[scale=0.33] at (4,3) {$(\!c_0,\!y_0y_{8}\!)$};
\node[scale=0.33] at (4.025,1.8) {$(\!\widehat u_4,\!y_4y_{12}\!)$};
\node[scale=0.33] at (4,2.85) {$(\!\overline{c_0},\!y_0y_{8}\!)$};
\node[scale=0.33] at (4.03,1.65) {$(\!\overline{\widehat u_4},\!y_4y_{12}\!)$};
\node[scale=0.33] at (4.025,2.4) {$(\!\widehat c_2,\!y_2y_{10}\!)$};
\node[scale=0.33] at (4.01,1.2) {$(\!0,\!y_6y_{14}\!)$};
\node[scale=0.33] at (4.025,2.25) {$(\!\overline{\widehat c_2},\!y_2y_{10}\!)$};
\node[scale=0.33] at (4.01,1.05) {$(\!1,\!y_6y_{14}\!)$};
\node[scale=0.33] at (4,2.7) {$(\!\widehat c_1,\!y_1y_{9}\!)$};
\node[scale=0.33] at (4.01,1.5) {$(\!0,\!y_5y_{13}\!)$};
\node[scale=0.33] at (4,2.55) {$(\!\overline{\widehat c_1} ,\!y_1y_{9}\!)$};
\node[scale=0.33] at (4.01,1.35) {$(\!1,\!y_5y_{13}\!)$};
\node[scale=0.33] at (4.025,2.1) {$(\!\widehat u_3,\!y_3y_{11}\!)$};
\node[scale=0.33] at (4.01,0.9) {$(\!0,\!y_7y_{15}\!)$};
\node[scale=0.33] at (4.025,1.95) {$(\!\overline{\widehat u_3},\!y_3y_{11}\!)$};
\node[scale=0.33] at (4.01,0.75) {$(\!1,\!y_7y_{15}\!)$};
\node[scale=0.33] at (4.025,0.6) {$c_0 = \tilde c_0 \oplus \widehat u_4$};
\node[scale=0.33] at (4.025,0.45) {$\overline{c_0} = c_0 \oplus 1$};
\node[scale=0.33] at (4.025,0.3) {$\overline{\widehat u_3}\!=\!\widehat u_3\!\oplus\!1$};
\end{scope}
\begin{scope}[shift={(-0.79,-2.6)}]
\node[scale=0.5] at (4,3.15) {$X_0$};
\node[scale=0.33] at (4,3) {$(\!\epsilon,\!y_{0}\!)$};
\node[scale=0.33] at (4,1.8) {$(\!\epsilon,\!y_{8}\!)$};
\node[scale=0.33] at (4,2.4) {$(\!\epsilon,\!y_{4}\!)$};
\node[scale=0.33] at (4.025,1.2) {$(\!\epsilon,\!y_{12}\!)$};
\node[scale=0.33] at (4,2.7) {$(\!\epsilon,\!y_{2}\!)$};
\node[scale=0.33] at (4.025,1.5) {$(\!\epsilon,\!y_{10}\!)$};
\node[scale=0.33] at (4,2.1) {$(\!\epsilon,\!y_{6}\!)$};
\node[scale=0.33] at (4.025,0.9) {$(\!\epsilon,\!y_{14}\!)$};
\node[scale=0.33] at (4,2.85) {$(\!\epsilon,\!y_{1}\!)$};
\node[scale=0.33] at (4,1.65) {$(\!\epsilon,\!y_{9}\!)$};
\node[scale=0.33] at (4,2.25) {$(\!\epsilon,\!y_{5}\!)$};
\node[scale=0.33] at (4.025,1.05) {$(\!\epsilon,\!y_{13}\!)$};
\node[scale=0.33] at (4,2.55) {$(\!\epsilon,\!y_{3}\!)$};
\node[scale=0.33] at (4.025,1.35) {$(\!\epsilon,\!y_{11}\!)$};
\node[scale=0.33] at (4,1.95) {$(\!\epsilon,\!y_{7}\!)$};
\node[scale=0.33] at (4.025,0.75) {$(\!\epsilon,\!y_{15}\!)$};
\end{scope}
\begin{scope}[]
\begin{scope}[shift={(-0.9,0)},scale=0.5,xscale=0.8]
\draw (1.75,0.25) -- (2.65,0.25);
\draw (1.75,-1.25) -- (2.65,-0.25);
\draw (1.75,-0.75) -- (2.65,-0.25);
\draw (1.75,-0.25) -- (2.65,0.25);
\draw (1.75,-1.75) -- (2.65,-0.75);
\draw (1.75,-3.25) -- (2.65,-1.25);
\draw (1.75,-2.75) -- (2.65,-1.25);
\draw (1.75,-2.25) -- (2.65,-0.75);
\end{scope}
\begin{scope}[scale=0.5, shift={(-2.77,0)}]
\draw (4.6,0.25) -- (5.22,0.25);
\draw (4.6,-0.25) -- (5.22,-0.25);
\draw (4.6,-0.75) -- (5.22,-0.25);
\draw (4.6,-1.25) -- (5.22,0.25);
\draw (4.53,-1.75) -- (5.25,-0.75);
\draw (4.53,-2.25) -- (5.25,-1.25);
\draw (4.53,-2.75) -- (5.25,-1.25);
\draw (4.53,-3.25) -- (5.25,-0.75);
\end{scope}
\begin{scope}[scale=0.5, shift={(-2.85,0)}]
\draw (7,0.25) -- (7.42,0.8);
\draw (7,-0.25) -- (7.42,0.5);
\draw (6.95,-0.75) -- (7.42,0.2);
\draw (6.95,-1.25) -- (7.42,-0.1);
\draw (7,-1.75) -- (7.42,-0.4);
\draw (7,-2.25) -- (7.42,-0.7);
\draw (7,-2.75) -- (7.42,-1);
\draw (7,-3.25) -- (7.42,-1.3);
\end{scope}
\begin{scope}[scale=0.5, shift={(-1.3,0)}]
\draw (6.8,0.8) -- (7.42,0.8);
\draw (6.8,0.2) -- (7.42,0.5);
\draw (6.8,0.5) -- (7.42,0.8);
\draw (6.8,-0.1) -- (7.42,0.5);
\draw (6.9,-0.4) -- (7.42,0.2);
\draw (6.9,-1) -- (7.42,-0.1);
\draw (6.9,-0.7) -- (7.42,0.2);
\draw (6.9,-1.3) -- (7.42,-0.1);
\draw (6.9,-1.6) -- (7.42,-0.4);
\draw (6.9,-1.9) -- (7.42,-0.4);
\draw (6.8,-2.2) -- (7.42,-0.7);
\draw (6.8,-2.5) -- (7.42,-0.7);
\draw (6.8,-2.8) -- (7.42,-1);
\draw (6.8,-3.1) -- (7.42,-1);
\draw (6.8,-3.4) -- (7.42,-1.3);
\draw (6.8,-3.7) -- (7.42,-1.3);
\end{scope}
\end{scope}
\begin{scope}[red,dash pattern=on 2pt off 0.5pt]
\begin{scope}[ shift={(-0.9,0)},scale=0.5,xscale=0.8]
\draw (1.75,0.25) -- (2.65,-1.75);
\draw (1.75,-1.25) -- (2.65,-2.25);
\draw (1.75,-0.75) -- (2.65,-1.75);
\draw (1.75,-0.25) -- (2.65,-2.25);
\draw (1.75,-1.75) -- (2.65,-2.75);
\draw (1.75,-3.25) -- (2.65,-3.25);
\draw (1.75,-2.75) -- (2.65,-2.75);
\draw (1.75,-2.25) -- (2.65,-3.25);
\end{scope}
\begin{scope}[scale=0.5, shift={(-2.77,0)}]
\draw (4.6,0.25) -- (5.25,-1.75);
\draw (4.6,-0.25) -- (5.25,-2.25);
\draw (4.6,-0.75) -- (5.25,-1.75);
\draw (4.6,-1.25) -- (5.25,-2.25);
\draw (4.53,-1.75) -- (5.23,-2.75);
\draw (4.53,-2.25) -- (5.23,-3.25);
\draw (4.53,-2.75) -- (5.23,-2.75);
\draw (4.53,-3.25) -- (5.23,-3.25);
\end{scope}
\begin{scope}[scale=0.5, shift={(-2.85,0)}]
\draw (7,0.25) -- (7.42,-1.6);
\draw (7,-0.25) -- (7.42,-1.9);
\draw (6.95,-0.75) -- (7.42,-2.2);
\draw (6.95,-1.25) -- (7.42,-2.5);
\draw (7,-1.75) -- (7.42,-2.8);
\draw (7,-2.25) -- (7.42,-3.1);
\draw (7,-2.75) -- (7.42,-3.4);
\draw (7,-3.25) -- (7.42,-3.7);
\end{scope}
\begin{scope}[scale=0.5, shift={(-1.3,0)}]
\draw (6.8,0.8) -- (7.42,-1.6);
\draw (6.8,0.2) -- (7.42,-1.9);
\draw (6.8,0.5) -- (7.42,-1.6);
\draw (6.8,-0.1) -- (7.42,-1.9);
\draw (6.9,-0.4) -- (7.42,-2.2);
\draw (6.9,-1) -- (7.42,-2.5);
\draw (6.9,-0.7) -- (7.42,-2.2);
\draw (6.9,-1.3) -- (7.42,-2.5);
\draw (6.9,-1.6) -- (7.42,-2.8);
\draw (6.9,-1.9) -- (7.42,-2.8);
\draw (6.8,-2.2) -- (7.42,-3);
\draw (6.8,-2.5) -- (7.42,-3.1);
\draw (6.8,-2.8) -- (7.42,-3.4);
\draw (6.8,-3.1) -- (7.42,-3.4);
\draw (6.8,-3.4) -- (7.42,-3.7);
\draw (6.8,-3.7) -- (7.42,-3.7);
\end{scope}
\end{scope}

\end{tikzpicture}
}
\caption{CSE of phase 5 of $K_{16}$ kernel}
\label{fCSEExample}
\end{figure}

\begin{example}
\label{eCSE}
Consider the identification of CSE for kernel $K_{16}$ at external phase 5 and internal phase $h_5 = 8$. The decoding window is $\mD_5 = \set{5,6,7}$. Thus, under condition $u_0^4 = \widehat u_0^4$ we have 
$\widehat \mZ_{5,8} = \set{v_0^7|\widehat u_0^4.v_5^7,v_i = \set{0,1}, i\in \mD_5 }$, which is initialized at line \ref{initZ} of Alg. \ref{fCSEAlg}.

Computation of $X_\lambda$, $0 \leq \lambda \leq 4,$  by 
\textit{GetCSEPairs}$(8,5)$ is illustrated in Fig. \ref{fCSEExample}. The set $X_4$ is initialized as $\set{(v_0^7|y_0^{15})|v_0^7\in \widehat \mZ_5^{(8)}}$ at line \ref{initX}. The elements of $X_3$ are obtained from elements of $X_4$ at lines \ref{r6}-\ref{r9}. 
Each pair $(v_0^7|y_0^{15}) \in X_4$ is connected with two distinct pairs $(\bar v_0^3|\bar y_0^{7}) \in X_3$. One of them is given by  \eqref{mRecA}  and connected by a black solid line, while another one is given by \eqref{mRecB} and connected by a red dashed line. This process is performed at lines \ref{r8}--\ref{r9}. The same process is performed to obtain the sets $X_2,X_1,X_0$. 

In a straightforward implementation of WP, for each value of 
$S_4^{(8)} \in \mathbb S_5^{(8)}$  one should compute $2$ values of $S_3^{(4)}$, $4$ values of $S_2^{(2)}$, and $8$ values of $S_1^{(1)}$. This results in $8\cdot(1+2+4+8)= 120$ operations. According to Fig. \ref{fCSEExample}, we have $|X_4|+|X_3|+|X_2|+|X_1| = 8\cdot3+16=40$, which implies that we can compute $\mathbb S_5^{(8)}$ only with 40 operations using  Alg. \ref{fComputeLLRsAlg}.
\end{example}

\subsection{Reusing of CSE at different phases}
\label{ssCSEReusing}
Arikan SC decoding allows one to reuse intermediate LLRs to compute LLR for the next phases as we mentioned in Section \ref{sLLRSimple}. Similar approach can be used in WP with CSE. 

Recall that in Arikan SC decoding at phase $i > 0$ one needs to recompute only the intermediate LLRs $S_{t - k}^{(\lfloor i/2^{s-k}\rfloor)}$, $0 \leq k \leq s$, where $s = \psi(i)$ is the largest integer such that $2^s$ divides $i$. Observe that $S_{t-s}^{(\lfloor i/2^{s}\rfloor)}$ is calculated via $P$ function \eqref{mMinSum2} using partial sums $C_{t-s}^{(\lfloor i/2^{s}\rfloor)}[\beta]$, while remaining $S_{t - k}^{(\lfloor i/2^{s-k}\rfloor)}$, $0 \leq k < s,$ are computed via $Q$ function \eqref{mMinSum1}.

The above principle is applicable to the case of WP with CSE as well. Consider processing of the $\phi$-th phase of  kernel $K$. Suppose we compute all CSE at the internal phase $h_{\phi}$. Let  $s' = \psi(h+1)$, and $|\mD_\phi| \geq |\mD_{\phi+1}|$. In this case, $X_\lambda^{(h_\phi+1)} \subseteq  X_\lambda^{(h_\phi)}$, $\lambda \in [s']$. Thus, at phase $h_\phi+1$  we need to recompute only CSE\ LLRs $S_\lambda^{(\lfloor h_\phi/2^{t-\lambda} \rfloor)}$ at layers $s' \leq \lambda \leq t$. In other words, one can start Alg. \ref{fComputeLLRsAlg} from layer $s'$. In contrast, in case of $|\mD_\phi| < |\mD_{\phi+1}|$ we have $X_\lambda^{(h_\phi)} \subseteq   X_\lambda^{(h_\phi+1)} $, $\lambda \in [t+1]$, and need to update all $X_\lambda$. However, even in this case we can reuse  $|X_\lambda^{(h_\phi+1)}|/|X_\lambda^{(h_\phi)}|$ LLRs at layers $\lambda \in [s']$.
\begin{example}
At external phase $\phi = 5$ of kernel $K'_{16}$ we have $u_5 = v_6\oplus v_9$, $h_5 = 9$, and $\mathcal D_5 = \mathcal D_4$. In this case, $X_\lambda^{(5)} =  X_\lambda^{(4)}, \lambda \in [4]$. Thus, one can start Alg. \ref{fComputeLLRsAlg}\ from layer 4. The set $\mathbb S_{4}^{(9)}$ is computed at lines \ref{lcEval}--\ref{lpEval} with $v_8$ evaluated as $\widehat u_4$.
\end{example}

  As a result, arithmetic complexity (number of summation and comparison operations) of computing $\mathbb S_\phi^{(h)}$ is given by $\sum_{\lambda = s'}^t |X_\lambda|$, while straightforward implementation of the WP
requires $|\mathcal D_{\phi}| \cdot \sum_{\lambda = \lambda'}^t 2^{t-\lambda}$ operations. It can be seen that 
$$\sum_{\lambda = \lambda '}^t |X_\lambda| \leq |\mathcal D_{\phi}| \cdot \sum_{\lambda = \lambda'}^t 2^{t-\lambda}$$
by definition of the CSE. 


\section{Further improvements of window processing}
\label{sWindSimpl}

In this section we  further simplify window processing by exploiting some properties  of maximization of path scores. 
\subsection{Path score computation}
\label{ssPathKernel}
We consider  WP of $l\times l$ polarization kernel $K$. Suppose for some external phase $\phi$ we have $h_\phi > h_{\phi-1} +\ 1$, or, in other words, $|\mathcal D_\phi|$ $>$ $|\mathcal D_{\phi-1}|$. As we discussed in Section \ref{s:Motivation}, in this case we  need to compute several input symbols LLRs 
$S_t^{(h)}$, where  $h_{\phi-1}<h < h_{\phi}$. 

Essentially, in this case WP requires to compute path scores
\begin{align}
\label{sScoreExt}
R_y(v_0^{h_\phi})=R_y(v_0^{h_{\phi}-1})+\tau\left(S_t^{(h_\phi)}, v_{h_\phi}\right) = 
\quad \quad \quad\nonumber \\
 R_y(v_0^{h_{\phi-1}})+
\sum_{h=h_{\phi-1}+1}^{h_\phi} \tau\left(S_t^{(h)}, v_{h}\right).
\end{align} 

There is a simple way to compute the last term in \eqref{sScoreExt}.
 Let 
$$E(c_0^{n-1},s_0^{n-1})=\sum_{i=0}^{n-1}\tau(s_i,c_i)$$
be an ellipsoidal weight 
(also known as correlation discrepancy) 
of vector $c_0^{n-1}\in \F_2^n$ with respect to  $s_0^{n-1}$   \cite{Valembois2004box,moorthy1997softdecision}. 

\begin{theorem}[\cite{trofimiuk2020fast}]
\label{tLogLikelihood} The ellipsoidal weight of the vector $v_0^{2^t-1}F_t$ with respect to the input LLRs $s$ is equal to the score of the path $v_0^{2^t-1}$ in the SC\ decoder, i.e. $$E(v_0^{2^t-1}F_t,s)=
\sum_{i=0}^{2^t-1}\tau(S_t^{(i)}(v_0^{i-1},y_0^{2^t-1}),v_i),$$
where $s=(S_{0}^{(0)}(y_0),\dots,S_{0}^{(0)}(y_{2^m-1})).$
\end{theorem}
Theorem \ref{tLogLikelihood} implies the following:
\begin{corollary}
\label{cScoreOuter}
Consider Arikan SC decoding of $(n = 2^t,k)$ polar code at phase $\phi = i+2^q-1$, where $q$ is the largest integer such that $2^q$ divides $i$.  Then,
$$
\sum_{\beta=i}^{i+2^q-1}\tau(S_t^{(\beta)}
(v_0^{\beta-1},y_0^{n-1})
,v_j) = \sum_{\beta=0}^{2^q-1}\tau(S_{\lambda-q}^{(\lfloor \phi/2^q \rfloor)}[\beta]
,c_\beta),
$$
where $c_0^{2^q-1} = v_i^{i+2^q-1}F_q$.
\end{corollary}

Corollary \ref{cScoreOuter} allows one to avoid the computation of some $S_t^{(h)}$ in \eqref{sScoreExt} and use $S_{t-q}^{(\lfloor h/2^q\rfloor)}[\beta]$ instead. In some cases, one can obtain $R(v_0^{h_\phi}|y_0^{n-1})$ in a lower number of operations. 

\begin{example}
\label{K16FFTExample}
Consider computing the LLR for $u_4$ of $K_{16}'$ kernel. The decoding window $\mD_4 = \set{3,5,6,7}$. According to \eqref{f:PathScore}, one has 
$$
R_y(v_0^8)=R_y(v_0^{7})+\tau\left(S_4^{(8)}(v_0^{7},y_0^{15}),v_8\right).
$$
By definition, we have
\begin{equation}
R_y(v_0^7) = \sum_{j=0}^{7}\tau(S_4^{(j)}
(v_0^{j-1},y_0^{15}),v_j).
\label{Rv7basic}
\end{equation}
Thus, to compute all $R_y(v_0^7)$, $v_0^7 \in \mZ_4,$ one needs to obtain intermediate LLRs $\mathbb S_4^{(5)},\mathbb S_4^{(6)}$ and $\mathbb S_4^{(7)}$ and sum corresponding $\tau(S_4^{(h)}, v_h)$. Using CSE and properties of $\tau$ function, this can be done with 
36 operations. 

Using Corollary \ref{cScoreOuter}, one can rewrite \eqref{Rv7basic} in the following way:
\begin{equation}
R_y(v_0^7) = R_y(v_0^3)+ 
 \sum_{\beta = 0}^{3}\tau(S_2^{(1)}[\beta],(v_4^7F_2)_\beta).
\label{Ry7c1}
\end{equation}
According to \eqref{fSExact}, $$S_2^{(1)}[\beta] =  S_2^{(1)}((v_0^3F_2)_\beta,
(y_\beta,y_{\beta+4},y_{\beta+8},y_{\beta+12})).$$ 
Note that $R_y(v_0^3)$ was computed at the previous phase. 
Unlike \eqref{sScoreExt}, one does not need to compute $\mathbb S_4^{(5)},\mathbb S_4^{(6)}$ and $\mathbb S_4^{(7)}$ in \eqref{Ry7c1}. To compute the last term of \eqref{Ry7c1} for all $v_0^7 \in \mZ_4$ we first observe that $\bar \mZ_b =\set{v_4^7F_2|v_4 = b, v_5^7 \in \bF_2^3}$ is a coset of Reed-Muller code $RM(1,2) \oplus b(F_2[0])$. Furthermore,  
\begin{equation}
\label{fFHT}
\sum_{\beta = 0}^{3}\tau(S_2^{(1)}[\beta],c_\beta )
\!= \!\frac{1}{2}\left(\sum_{\beta=0}^3(-1)^{c_\beta}S_2^{(1)}[\beta]-\!\sum_{\beta = 0}^3 |S_2^{(1)}[\beta]|\!\right),
\end{equation}
where $c_0^3=v_4^7F_2$. Assume for simplicity that $v_4=0$. Then, using the fast Hadamard transform (FHT) \cite{beery1986optimal}, one can obtain the term
$\sum_{\beta=0}^3(-1)^{c_\beta}S_2^{(1)}[\beta]$ for all $v_4^7\in {\bar \mZ_0}$ simultaneously  in 8 operations. The term 
$\sum_{\beta = 0}^3 |S_2^{(1)}[\beta]|$ does not need to be computed, since it is constant and cancels in \eqref{mKernLLR}. 

Recall that we have the index $3 \in \mD_4$, so, we have two arrays $S_2^{(1)}[\beta]$ of intermediate LLRs computed for $v_3 \in \set{0,1}$. Fortunately, these  LLRs were already obtained at phase 3, since $\mD_3 = \set{3}$. Thus, to compute $R_y(v_0^7)$ for all $v_0^7$ $\in$ $\mZ_4$ we need to evaluate \eqref{fFHT} twice for different $S_2^{(1)}[\beta]$ and add the results to $R_y(v_0^3)$. We can also set $R_y(v_0^2) = 0$, since $v_0^2$ is fixed, so, only 8 values of $R_y(v_0^3)$ is nonzero, see Section \ref{ssLLRWind}.
As a result,  Corollary \ref{cScoreOuter} and FHT allow us to reduce the path score computation from 36 to 24 operations.
\end{example}
\subsection{Recursive maximum computation}
\label{ssRecMax}
Consider the case when the decoding window is monotonically decreasing at the next phases of window processing. In this case, the complexity of \eqref{mKernLLR} can be drastically reduced. 

Let $h_\phi = h_{\phi+1} = \cdots = h_{\phi+q}$ at some phase $\phi$. It implies that $|\mD_{\phi+i}| = |\mD_{\phi}|-i, i \in [q+1],$ and, therefore 
$\mathcal Z_{\phi+q} \subset \mathcal Z_{\phi+q - 1} \subset \cdots \subset \mathcal Z_{\phi}.$ 
%
%
Thus, each path score $R_y(v_0^{h_{\phi+i}})$ for $v_0^{h_{\phi+i}} \in \mZ_{\phi+i}$ in fact is computed at the phase $\phi$. Recall that WP \eqref{mKernLLR} for  phase $\phi+i$ requires the maximization of $R_y(v_0^{h_{\phi+i}})$ over ${v_0^{h_{\phi+i}}\in\mathcal Z_{\phi+1}^{(b)}}$, $b \in \bF_2$. Since $h_{\phi+i} = h_{\phi}$ and $\mathcal Z_{\phi+i} \subset \mathcal Z_{\phi}$, we propose to perform this maximization during the computation of $\bS_{1,\phi}$.  To do this, it is convenient to define the array 
\begin{equation}
\label{fMEq}
M_{i}[\bar u_\phi^{\phi+i-1}.b] = \max_{\substack{v_0^{h_{\phi+i}}\in\mathcal Z_{\phi+i}^{(b)}, \quad u_{\phi}^{\phi+i-1}=  \bar u_\phi^{\phi+i-1}}} R_y(v_0^{h_{\phi}}),
\end{equation}
where $i \in [q+1]$.  This array can be  defined recursively as 
\begin{equation}
\label{fMaxReq}
M_{i}[\bar u_\phi^{\phi+i-1}.b] = \max(M_{i+1}[\bar u_\phi^{\phi+i-1}.b.0],M_{i+1}[\bar u_\phi^{\phi+i-1}.b.1]).
\end{equation}

 The base of this recursion is given by $M_q[\bar u_\phi^{\phi+q-1}.b], b \in \bF_2$, which should be computed by \eqref{fMEq}. As a result, after recursive computation of $M_0[b], b\in \bF_2,$ according to \eqref{mKernLLR}, the $(\phi + i)$-th LLR $\bS_{1,\phi+i}$ can be obtained  in one subtraction as $M_{i}[\widehat u_\phi^{\phi+i-1}.0]-M_{i}[\widehat u_\phi^{\phi+i-1}.1]$, where symbols $\widehat u_\phi^{\phi+i-1}$ are estimated by the decoder. Observe that the number of comparisons to obtain $M_0[b]$ by \eqref{fMaxReq} is the same as for straightforward maximization
$\max_{v_0^{h_{\phi}}\in\mathcal Z_{\phi}^{(b)}} R_y(v_0^{h_\phi})$ in \eqref{mKernLLR}, we just propose to find this maximum in such way that intermediate comparisons can be used at next $q$ phases.

\begin{example}
\label{K16RecMax}
Consider external phase 7 of $K_{16}$ kernel processing. For this phase, $h_7 =h_{8}= h_{9} = h_{10} = 10$, $\mD_7 = \set{5,6,7}$, thus, $\mD_8 = \mD_7 \setminus \set{5} = \set{6,7}$, $\mD_9 = \mD_8 \setminus \set{6} = \set{7}$, and $\mD_{10} = \mD_9 \setminus \set{7} = \set{}$. It implies that the recursive maximum computation can be used  for phases 8--10. 

Recall that $u_7 = v_5 \oplus v_6 \oplus v_{10}$, $u_8 = v_5$, $u_9 = v_6$, and $u_{10} = v_7$. Here $q = 3$, so, to compute $M_0[b], b \in \bF_2$, we start from $M_3[(v_5 \oplus v_6 \oplus v_{10}).v_5.v_6.v_7] = R_y(v_0^{10})$, for $v_0^{10} \in \mZ_7$. One step of \eqref{fMaxReq} results in computing $M_2[(v_5 \oplus v_6 \oplus v_{10}).v_5.v_6]$, which are equal to $\max_{v_7 \in \bF_2} R_0(v_0^{10})$ for some fixed $v_5,v_6,v_{10}$. Note that the term $\max_{v_7 \in \bF_2} R_0(v_0^{10})$ is a part of in WP \eqref{mKernLLR} for phase 9.  After computing $M_0[b]$ one immediately obtain $\bS_{1,7} = M_0[0] - M_0[1]$. At the next phase, using value $\widehat u_7$, provided by the decoder, we have $\bS_{1,8} = M_1[\widehat u_7.0] - M_1[\widehat u_7.1]$.
\end{example}

\subsection{Reusing of maximums from the previous phase}
\label{ssMaxReusing}

Expression \eqref{mKernLLR} requires computing of two maximums over the path scores $R_y(v_0^{h_{\phi}})$ for $v_0^{h_{\phi}} \in \mZ_{\phi}^{(b)}$, $b \in \bF_2$. It is possible to reduce the complexity of this step.
Namely, consider the case $h_{\phi+1} = h_{\phi}+1$, which implies that $\mD_{\phi+1} = \mD_{\phi}$. According to \eqref{f:PathScore}, the path score 
$$R_y(v_0^{h_{\phi}+1}) = R_y(v_0^{h_\phi})+
\tau(S_t^{(h_{\phi}+1)}(v_0^{h_{\phi}},y_0^{l-1}), v_{h_{\phi}+1}).$$ Recall that $\tau(S,c)$ function is zero for one of $c \in \bF_2$, thus, the path score $R_y(v_0^{h_{\phi}+1}) = R_y(v_0^{h_\phi})$ for some $v_{h_{\phi}+1}$. 

Suppose the decoder makes an estimate of the symbol $\widehat u_\phi$. Let 
$\bar v_0^{h_\phi} = \arg \max_{v_0^{h_\phi} \in \mZ_{\phi}^{(\widehat u_\phi)}} R_y(v_0^{h_\phi}).$
Let $b \in \bF_2$ be such that  $\tau(S_t^{(h_{\phi}+1)}(\bar v_0^{h_{\phi}},y_0^{l-1}),b) = 0$. Using \eqref{fTransformMinSpanV} for $v_{h_\phi+1}$ with $u_{\phi+1} = b$ we can obtain such value $a \in \bF_2$ so that vector $\bar v_0^{h_{\phi}}.a \in \mZ_{\phi+1}^{(b)}$. Therefore, $R_y(\bar v_0^{h_\phi}.a) = R_y(\bar v_0^{h_\phi})$ and $\bar v_0^{h_\phi}.a = \arg \max_{v_0^{h_{\phi+1}} \in \mZ_{\phi+1}^{(b)}} R_y(v_0^{h_{\phi}+1})$. It means that by tracking the value of $b$ we can directly obtain $\max_{v_0^{h_{\phi+1}} \in \mZ_{\phi+1}^{(b)}} R_y(v_0^{h_{\phi}+1})$ in  \eqref{mKernLLR}. It remains to compute only  $\max_{v_0^{h_{\phi+1}} \in \mZ_{\phi+1}^{(b \oplus 1)}} R_y(v_0^{h_{\phi}+1})$.   

This approach can be employed in the case of reduced decoding window at the next phases.

\section{Efficient processing of some  kernels}
\label{sEfficientProc}
In this section we briefly describe the window processing of some polarization kernels with good polarization properties and small size of decoding windows. We carefully compute the arithmetic complexity, and point out all simplifications used. Observe that the proposed algorithm involves addition and comparison operations only.

\subsection{Overall processing algorithm}

\begin{algorithm}
\caption{Improved window processing algorithm for phase $\phi$ of $2^t \times 2^t$ kernel $K$} 
\begin{enumerate}
\item[Step 1]: If $h_\phi > h_{\phi-1} +\ 1$, then use Corollary \ref{cScoreOuter} to obtain $R_y(v_0^{h_\phi -1}), v_0^{h_\phi -1} \in \mZ_\phi$.\

\item[Step 2]: Compute $S_t^{(h_\phi)}(v_0^{h_\phi -1},y_0^{l-1}),v_0^{h_\phi -1} \in \mZ_\phi$ using 

CSE via Alg. \ref{fComputeLLRsAlg}. 
Reuse the intermediate LLRs from 

the previous phases if possible (see Section \ref{ssCSEReusing}).

\item[Step 3]: Compute $R_y(v_0^{h_\phi})$ for $v_0^{h_\phi} \in \mZ_\phi$ with $2^{|\mD_\phi|}$ operations (see Section \ref{ssLLRWind}).

\item[Step 4]:\ \label{WPMax} Compute $\mathcal R_b = \max_{v_0^{h_{\phi}}\in\mathcal Z_{\phi}^{(b)}} R_y(v_0^{h_\phi})$ for $b \in [2]$ with $2^{|\mD_\phi|+1} - 2$ operations. Some
modifications are possible:
\begin{enumerate}
\item \label{fMaxReuse} If $h_{\phi} = h_{\phi-1}+1$, then reuse computations 

from previous phase to obtain $\mathcal R_b$  and compute only $\mathcal R_{b \oplus 1}$ with $2^{|\mD_\phi|} -1$ operations 

(see Section \ref{ssMaxReusing}). 

\item \label{fAM} If $h_\phi = h_{\phi+1} = \cdots = h_{\phi+q}$, then compute $\mR_b$ 

as $M_0[b]$ using \eqref{fMaxReq} 
with 
$2^{|\mD_\phi|+1} - 2$ 

operations.
\end{enumerate}
\item[Step 5]:\ $\bS_{1,\phi} \gets \mR_0- \mR_1$.
\end{enumerate}

%
%
%
%
%
%
%
\label{fWPAlg}
\end{algorithm}

 We consider the processing of a $2^t \times 2^t$ polarization kernel $K$. Alg. \ref{fWPAlg} illustrates all stages of WP of phase $\phi$ with  improvements proposed in the previous sections. Observe that the transition matrix $T$, equation \eqref{fTransformMinSpanV}, and CSE pairs (see Section \ref{ssCSEIdent}) should be computed offline.

Observe that Alg. \ref{fWPAlg} can be skipped for some phases $\phi$. If $\mD_\phi = \emptyset$, then WP reduces to Arikan SC decoding, i.e. $\bS_{1,\phi} = S_t^{h_\phi}$. In the case 
$h_{\phi-p} = \cdots = h_\phi = \cdots =h_{\phi+q}$  we assume that line 
4b of Alg. \ref{fWPAlg} is performed at phase $h_{\phi-p}$, hence 
$\bS_{1,\phi} = M_{p}[\widehat u_{\phi-p}^{\phi-1}0]-M_{p}[\widehat u_{\phi-p}^{\phi-1}1]$
(Section \ref{ssRecMax}).

The memory requirements for this algorithm are following. The total number of CSE stored is given by $C_s = \sum_{\lambda = 0} ^t \max_{\phi \in [l]} |X_\lambda^{( h_\phi )}|$ values. We also need to store the path scores, which requires $C_r = \max_{\phi \in [l]} 2^{|\mD_\phi|+1}$ values in the worst case. If recursive maximum computation (Section \ref{ssRecMax}) is applicable, then we additionally need 
$C_m = \max_{\phi \in [l]: h_\phi  = \cdots = h_{\phi+q}} (2^{q+1}-2)$ LLR entries.
Therefore, the overall memory requirements are $C_s+C_r+C_m$ LLR entries. 

\subsection{$K_{16}$ kernel}
\label{ssDescr16_2}
At phases 0--4, 11--15, $\mD_\phi = \emptyset$, thus, WP is equal to Arikan SC decoding. The processing complexity for these phases is given in Table \ref{tDecWin16}. Observe that all intermediate LLRs for phases 11--15 can be retrieved from CSE LLRs computed at the previous phases.

For phases 8--10 one can use the recursive maximum computation, as shown in Example \ref{K16RecMax}.

Table \ref{K16Compl} demonstrates the complexity of all Alg. \ref{fWPAlg} steps for phases 5--7, including CSE sizes. The complexity figures in this table take into account reusing of CSE. We also indicate the type of simplification implemented at step 4. 

At step 1 of phase 5, the path scores $R_y(v_0^7)$ for $v_0^7 \in \mZ_5$  can be computed with 8 
operations (see Section \ref{ssPathKernel}). Moreover, using the properties of FHT, one can find $\arg \max_{v_0^7 \in \mZ_5} R_y(v_0^7)$ with 3 operations and obtain one of $\mR_b$. The remaining $\mR_{b \oplus 1}$ is computed with 7 operations.

\begin{table}[t]
\caption{Complexity of Alg. \ref{fWPAlg} for $K_{16}$ kernel}
\centering
\scalebox{1}{
\small
\begin{tabular}{|l|l|l||l|l|l|l|l|l|c|l|c|}
\hline
\multirow{3}{*}{$\phi$} & \multirow{3}{*}{$h_\phi$} & \multirow{3}{*}{$|\mD_\phi|$} & \multicolumn{8}{c|}{Complexity of Alg. \ref{fWPAlg} steps}                                                   & \multirow{3}{*}{\begin{tabular}[c]{@{}c@{}}Total\\ complexity\end{tabular}} \\ \cline{4-11}
                        &                           &                               & \multirow{2}{*}{1} & \multicolumn{4}{c|}{2}        & \multirow{2}{*}{3} & \multirow{2}{*}{4} & \multirow{2}{*}{5} &                                                                             \\ \cline{5-8}
                        &                           &                               &                    & $X_1$ & $X_2$ & $X_3$ & $X_4$ &                    &                    &                    &                                                                             \\ \hline
5                       & 8                         & 3                             & 8                  & 16    & 8     & 8     & 8     & 8                  & 10                 & 1                  & 67                                                                          \\ \hline
6                       & 9                         & 3                             & -                  & -     & -     & -     & 8     & 8                  & 7 (4a)             & 1                  & 24                                                                          \\ \hline
7                       & 10                        & 3                             & -                  & -     & -     & 16    & 8     & 8                  & 14 (4b)            & 1                  & 47                                                                          \\ \hline
\end{tabular}}
\label{K16Compl}
\end{table}

The total processing complexity of $K_{16}$ is 181 operations, including 95 additions and 86 comparisons.
\subsection{$K_{32}$ kernel}
\begin{figure}[ht]
\centering
\scalebox{0.9}{
\footnotesize
\parbox{0.5\textwidth}
{ 
$
\arraycolsep=1.35pt\def\arraystretch{0.7}
\begin{array}{c}
K_{32}, E = 0.521936, \mu = 3.417\\
{\left(
\begin{array}{cccccccccccccccccccccccccccccccc}
1&0&0&0&0&0&0&0&0&0&0&0&0&0&0&0&0&0&0&0&0&0&0&0&0&0&0&0&0&0&0&0\\
1&1&0&0&0&0&0&0&0&0&0&0&0&0&0&0&0&0&0&0&0&0&0&0&0&0&0&0&0&0&0&0\\
1&0&1&0&0&0&0&0&0&0&0&0&0&0&0&0&0&0&0&0&0&0&0&0&0&0&0&0&0&0&0&0\\
1&1&1&1&0&0&0&0&0&0&0&0&0&0&0&0&0&0&0&0&0&0&0&0&0&0&0&0&0&0&0&0\\
1&0&0&0&1&0&0&0&0&0&0&0&0&0&0&0&0&0&0&0&0&0&0&0&0&0&0&0&0&0&0&0\\
1&0&0&0&0&0&0&0&1&0&0&0&0&0&0&0&0&0&0&0&0&0&0&0&0&0&0&0&0&0&0&0\\
1&1&0&0&0&0&0&0&1&1&0&0&0&0&0&0&0&0&0&0&0&0&0&0&0&0&0&0&0&0&0&0\\
1&0&1&0&0&0&0&0&1&0&1&0&0&0&0&0&0&0&0&0&0&0&0&0&0&0&0&0&0&0&0&0\\
1&0&1&0&1&1&0&0&0&1&1&0&0&0&0&0&0&0&0&0&0&0&0&0&0&0&0&0&0&0&0&0\\
0&1&1&0&1&0&1&0&1&1&0&0&0&0&0&0&0&0&0&0&0&0&0&0&0&0&0&0&0&0&0&0\\
1&1&1&1&1&1&1&1&0&0&0&0&0&0&0&0&0&0&0&0&0&0&0&0&0&0&0&0&0&0&0&0\\
1&1&1&1&0&0&0&0&1&1&1&1&0&0&0&0&0&0&0&0&0&0&0&0&0&0&0&0&0&0&0&0\\
1&0&0&0&0&0&0&0&0&0&0&0&0&0&0&0&1&0&0&0&0&0&0&0&0&0&0&0&0&0&0&0\\
1&1&0&0&0&0&0&0&0&0&0&0&0&0&0&0&1&1&0&0&0&0&0&0&0&0&0&0&0&0&0&0\\
0&1&0&0&1&0&0&0&1&0&0&0&1&0&0&0&1&1&0&0&0&0&0&0&0&0&0&0&0&0&0&0\\
1&1&0&0&1&1&0&0&1&1&0&0&1&1&0&0&0&0&0&0&0&0&0&0&0&0&0&0&0&0&0&0\\
1&0&1&0&0&0&0&0&0&0&0&0&0&0&0&0&1&0&1&0&0&0&0&0&0&0&0&0&0&0&0&0\\
1&1&1&1&0&0&0&0&0&0&0&0&0&0&0&0&1&1&1&1&0&0&0&0&0&0&0&0&0&0&0&0\\
0&1&0&1&1&0&1&0&1&0&1&0&1&0&1&0&1&1&1&1&0&0&0&0&0&0&0&0&0&0&0&0\\
1&1&1&1&1&1&1&1&1&1&1&1&1&1&1&1&0&0&0&0&0&0&0&0&0&0&0&0&0&0&0&0\\
1&0&0&0&1&0&0&0&0&0&0&0&0&0&0&0&1&0&0&0&1&0&0&0&0&0&0&0&0&0&0&0\\
1&0&0&0&0&0&0&0&1&0&0&0&0&0&0&0&1&0&0&0&0&0&0&0&1&0&0&0&0&0&0&0\\
1&1&0&0&0&0&0&0&1&1&0&0&0&0&0&0&1&1&0&0&0&0&0&0&1&1&0&0&0&0&0&0\\
1&0&1&0&0&0&0&0&1&0&1&0&0&0&0&0&1&0&1&0&0&0&0&0&1&0&1&0&0&0&0&0\\
1&0&1&0&1&1&0&0&0&1&1&0&0&0&0&0&1&0&1&0&1&1&0&0&0&1&1&0&0&0&0&0\\
0&1&1&0&1&0&1&0&1&1&0&0&0&0&0&0&0&1&1&0&1&0&1&0&1&1&0&0&0&0&0&0\\
1&1&1&1&1&1&1&1&0&0&0&0&0&0&0&0&1&1&1&1&1&1&1&1&0&0&0&0&0&0&0&0\\
1&1&1&1&0&0&0&0&1&1&1&1&0&0&0&0&1&1&1&1&0&0&0&0&1&1&1&1&0&0&0&0\\
1&0&0&0&1&0&0&0&1&0&0&0&1&0&0&0&1&0&0&0&1&0&0&0&1&0&0&0&1&0&0&0\\
1&1&0&0&1&1&0&0&1&1&0&0&1&1&0&0&1&1&0&0&1&1&0&0&1&1&0&0&1&1&0&0\\
1&0&1&0&1&0&1&0&1&0&1&0&1&0&1&0&1&0&1&0&1&0&1&0&1&0&1&0&1&0&1&0\\
1&1&1&1&1&1&1&1&1&1&1&1&1&1&1&1&1&1&1&1&1&1&1&1&1&1&1&1&1&1&1&1\\
\end{array}
\right)}
\end{array}
$
}}
\caption{$32 \times 32$ kernel with improved polarization}
\label{f:Kernel32}
\end{figure}

\begin{table}[ht]
\caption{Transition matrix of $K_{32}$ kernel}
\label{tDecWin32}
\centering
\scalebox{0.85}{
\footnotesize
\begin{tabular}{|l|p{0.09\textwidth}
|@{\hspace{0.01mm}}>{\centering}p{19mm}@{\hspace{0.01mm}}
|c||c|p{0.11\textwidth}
|@{\hspace{0.01mm}}>{\centering}p{15mm}@{\hspace{0.01mm}}
|c|}
\hline
$\phi$ &  $u_\phi$ & $\mD_\phi$ & Cost &$\phi$ &$u_\phi$ & $\mD_\phi$&Cost \\ \hline
0  & $v_0$              & $\set{}$ &31& 16&$v_{18}$                      &$\set{14,15}$&16\\ \hline
1  & $v_1$              & $\set{}$ &1& 17 &$v_{14}\oplus v_{19}$                      &$\set{14,15}$  &15\\ \hline
2  & $v_2$              & $\set{}$ &3& 18& $v_{14}$                      & $\set{15}$ &1\\ \hline
3  & $v_3$              & $\set{}$ &1& 19&$v_{15}$                     &$\set{}$  &1\\ \hline
4  & $v_4$              & $\set{}$ &7& 20 & $v_{20}$               &$\set{}$  &7\\ \hline
5  & $v_8$      & $\set{5,6,7}$ &67& 21&$v_{24}$               & $\set{21,22,23}$ &67\\ \hline
6  & $v_5\oplus v_6 \oplus v_{9} $   & $\set{5,6,7}$ &24& 22 &
$v_{21}\oplus v_{22} \oplus v_{25}$       & $\set{21,22,23}$ &24\\ \hline
7  & $v_5 \oplus v_{10}$              & $\set{5,6,7}$ &47& 23 & 
$v_{21}\oplus v_{26}$      & $\set{21,22,23}$ &47\\ \hline
8  & $v_{5}$           & $\set{6,7}$ &1& 24 &$v_{21}$              & $\set{22,23}$ &1\\ \hline
9  & $v_{6}$           & $\set{7}$ &1& 25 &$v_{22}$             & $\set{23}$ &1\\ \hline
10 & $v_{7}$              & $\set{}$ &1& 26 & $v_{23}$                  & $\set{}$ &1\\ \hline
11 & $v_{11}$              & $\set{}$ &1& 27 & $v_{27}$                 & $\set{}$ &1\\ \hline
12 & $v_{16}$           & $\set{12,13,14,15}$ &127& 28 &$v_{28}$                   & $\set{}$ &7\\ \hline
13 & $v_{12} \oplus v_{17}$           & $\set{12,13,14,15}$ &63& 29& $v_{29}$                   & $\set{}$ &1\\ \hline
14 & $v_{12}$           & $\set{13,14,15}$ &1& 30&$v_{30}$                   & $\set{}$ &3\\ \hline
15 & $v_{13}$           & $\set{14,15}$ &1& 31&$v_{31}$                   & $\set{}$ &1\\ \hline
\end{tabular}}
\end{table}

We present a new $32\times 32$ polarization kernel $K_{32}$, shown in Fig. \ref{f:Kernel32}. This kernel has rate of polarization $E(K_{32}) = 0.521936$ and scaling exponent $\mu(K_{32}) = 3.417$. We constructed this kernel via  the algorithm presented in \cite{trofimiuk2019construction32}. Table \ref{tDecWin32} demonstrates the transition matrix, decoding windows and processing complexity for each phase.

At phases  0--4, 11, 20, 27--31 $\mD_\phi = \emptyset$, the processing complexity is equal to that of Arikan SC decoding. Surprisingly, phases 5-10 and 21-26 of $K_{32}$ have absolutely the same structure and processing complexity as phases 5-10 of $K_{16}$ (see Section \ref{ssDescr16_2}). The only difference is in the expressions for $v_h$ and in the specific CSE pairs. However, the sizes of CSE sets are the same.

\begin{table}[]
\caption{Complexity of Algorithm \ref{fWPAlg} for $K_{32}$ kernel}
\centering
\scalebox{0.896}{
\small
\begin{tabular}{|l|l|l||l|l|l|l|l|l|l|c|l|c|}
\hline
\multirow{3}{*}{$\phi$} & \multirow{3}{*}{$h_\phi$} & \multirow{3}{*}{$|\mD_\phi|$} & \multicolumn{9}{c|}{Complexity of Algorithm \ref{fWPAlg} steps}                                                           & \multirow{3}{*}
                                                   {\begin{tabular}[c]{@{}c@{}}Total\\ complexity\end{tabular}} \\ \cline{4-12}
                        &                           &                               & \multirow{2}{*}{1} & \multicolumn{5}{c|}{2}                & \multirow{2}{*}{3} & \multirow{2}{*}{4} & \multirow{2}{*}{5} &                                                                             \\ \cline{5-9}
                        &                           &                               &                    & $X_1$ & $X_2$ & $X_3$ & $X_4$ & $X_5$ &                    &                    &                    &                                                                             \\ \hline
12                      & 16                        & 4                             & 15                 & 32    & 16    & 8     & 8     & 16    & 16                 & 15                 & 1                  & 127                                                                         \\ \hline
13                      & 17                        & 4                             & -                  & -     & -     & -     & -     & 16    & 16                 & 30 (4b)            & 1                  & 63                                                                          \\ \hline
16                      & 18                        & 2                             & -                  & -     & -     & -     & 4     & 4     & 4                  & 3 (4a)             & 1                  & 16                                                                          \\ \hline
17                      & 19                        & 2                             & -                  & -     & -     & -     & -     & 4     & 4                  & 6 (4b)             & 1                  & 15                                                                          \\ \hline
\end{tabular}
}
\label{K32Compl}
\end{table}

Table \ref{K32Compl} demonstrates the complexity of all steps of Alg. \ref{fWPAlg} for phases 12--13 and 16--17, including the sizes of CSE. At phase 12, we have
\begin{equation}
\label{fphase12}
R_y(v_0^{15}) = R_y(v_0^{12}) +\sum_{\beta = 0}^3 \tau(S_{3}^{(3)}[\beta],c_\beta),
\end{equation}where $c_0^3 = v_{12}^{15}F_2$ (see Corollary \ref{cScoreOuter}). The structure of $\mD_{12}$ implies that $c_0^3 \in \bF_2^4$. The equation \eqref{fphase12} requires 4 values $S_3^{(3)}[\beta]$. After that, using the properties of $\tau$ function, the last part of \eqref{fphase12} can be computed with 11 operations, while $R_y(v_0^{12})$ can be omitted. Moreover, since $c_0^3 = v_{12}^{15}F_2$, there is $\bar c_0^3$ such that $\sum_{\beta = 0}^3 \tau(S_{4}^{(3)}[\beta],\bar c_\beta) = 0$, thus, $\arg \max_{v_0^{15} \in \mZ_{15}} R_y(v_0^{12})$ can be immediately obtained with 0 operations. Thus, the total processing complexity for $K_{32}$ is 571 operations, including 297 additions and 274 comparisons.

\section{Numeric Results}
\label{sNumberic}
The performance of all codes was investigated for the case of AWGN\ channel with BPSK\ modulation. The sets of frozen symbols were obtained by the method \cite{trifonov2019construction}.

Observe that the processing techniques described in the previous sections can be also used to implement SCL\ decoder for polar codes with the considered kernels, using a straightforward generalization of the algorithm and data structures presented in \cite{tal2015list}.
The  SCL algorithm was implemented using the randomized order statistic algorithm  for the selection of the paths to be killed at each phase, which has the complexity $O(L)$, where $L$ is a list size.


\begin{figure}
\centering
\begin{subfigure}[b]{0.49\textwidth}
\includegraphics[width=\linewidth]{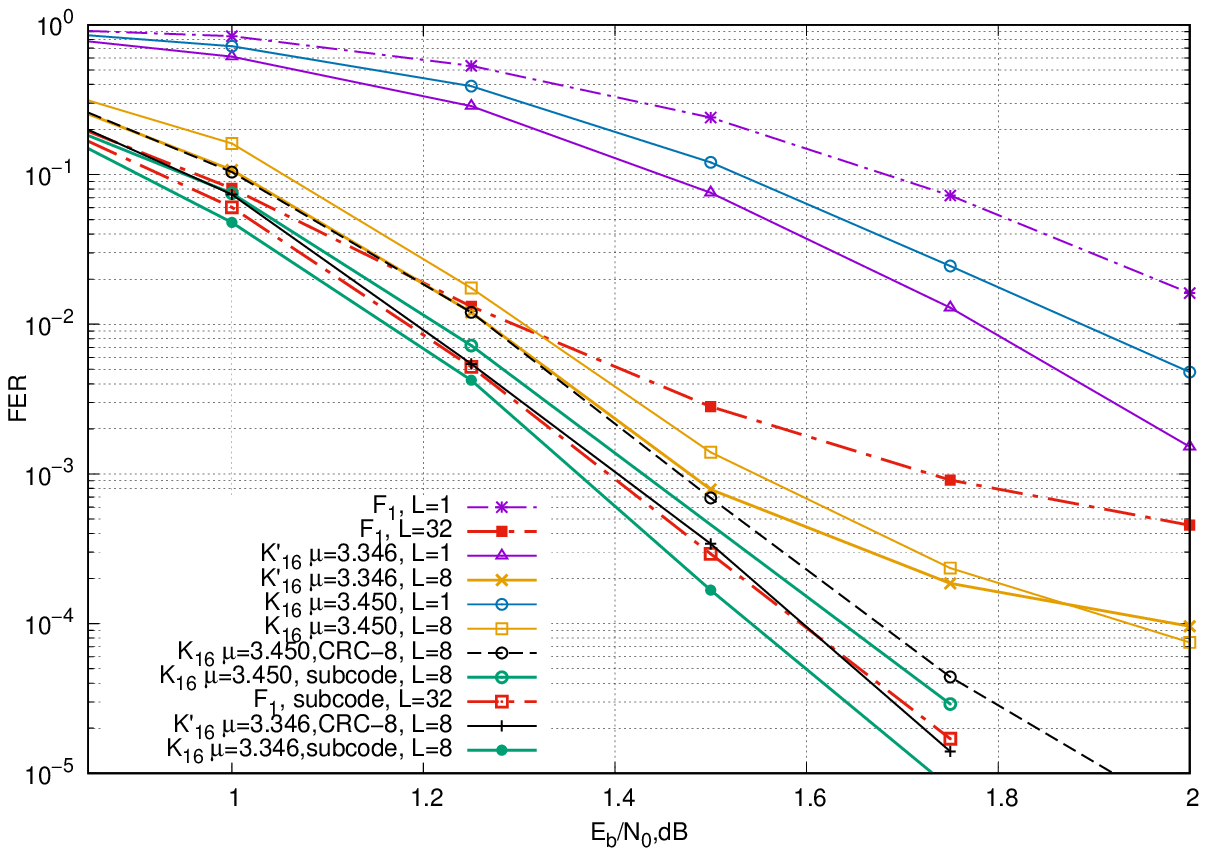}
\caption{$(4096, 2048)$ polar  codes}
\label{f4096_2048}
\end{subfigure}
\begin{subfigure}[b]{0.49\textwidth}
\includegraphics[width=\linewidth]{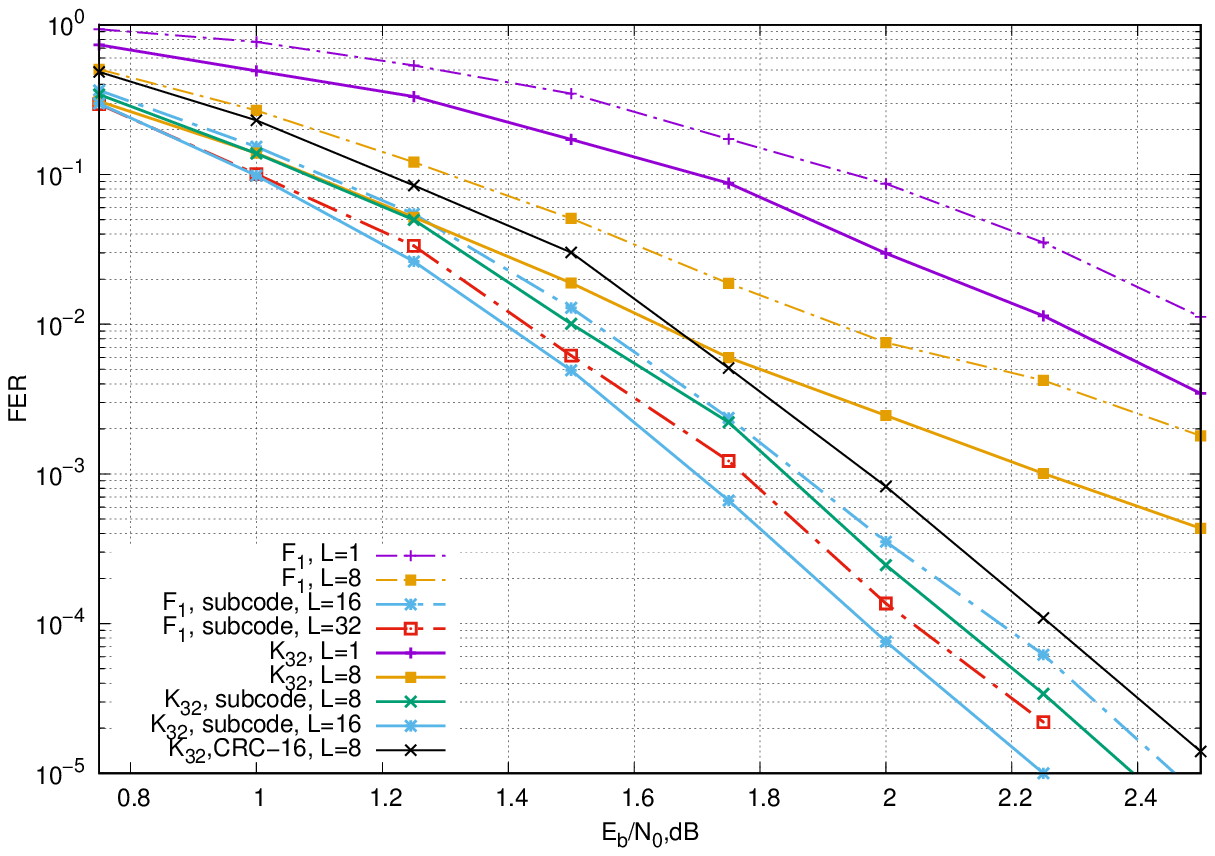}
\caption{$(1024, 512)$ polar  codes}
\label{f1024_512}
\end{subfigure}
\caption{Performance of polar codes with the proposed kernels}
\end{figure}



\subsection{Comparison with Arikan kernel} 

We constructed $(4096,2048)$ and $(1024,512)$ polar codes with kernels $K'_{16}, K_{16}$ and $K_{32}$ respectively.
Fig. \ref{f4096_2048} illustrates the  performance of $(4096,2048)$ plain polar codes, polar codes with CRC and polar subcodes (PSCs)  \cite{trifonov2017randomized}. The same polar code constructions were used for $(1024,512)$ polar codes, whose performance is demonstrated in Fig. \ref{f1024_512}. It can be seen that the codes based on kernels $K'_{16}, K_{16}$ and $K_{32}$ with improved polarization rate ($E(K'_{16}) = E(K_{16}) = 0.51828$, $E(K_{32}) = 0.521936$) provide significant performance gain compared with polar codes with Arikan kernel $F_1$.  Observe also that randomized PSCs provide better performance compared with polar codes with CRC.  
PSCs with kernels $K'_{16}, K_{16}$ under SCL with $L=8$ have almost the same performance as PSCs with $F_1$ kernel under SCL with $L = 32$.  Observe that the codes based on  kernels with lower scaling exponent exhibit better performance. In the case of PSCs with kernel $K_{32}$ their performance under SCL with $L=8$ is slightly better than the performance of PSC with $F_1$ kernel under SCL with $L=16$.  

%

\begin{figure}
\begin{subfigure}[b]{0.2415\textwidth}
\includegraphics[width=\linewidth]{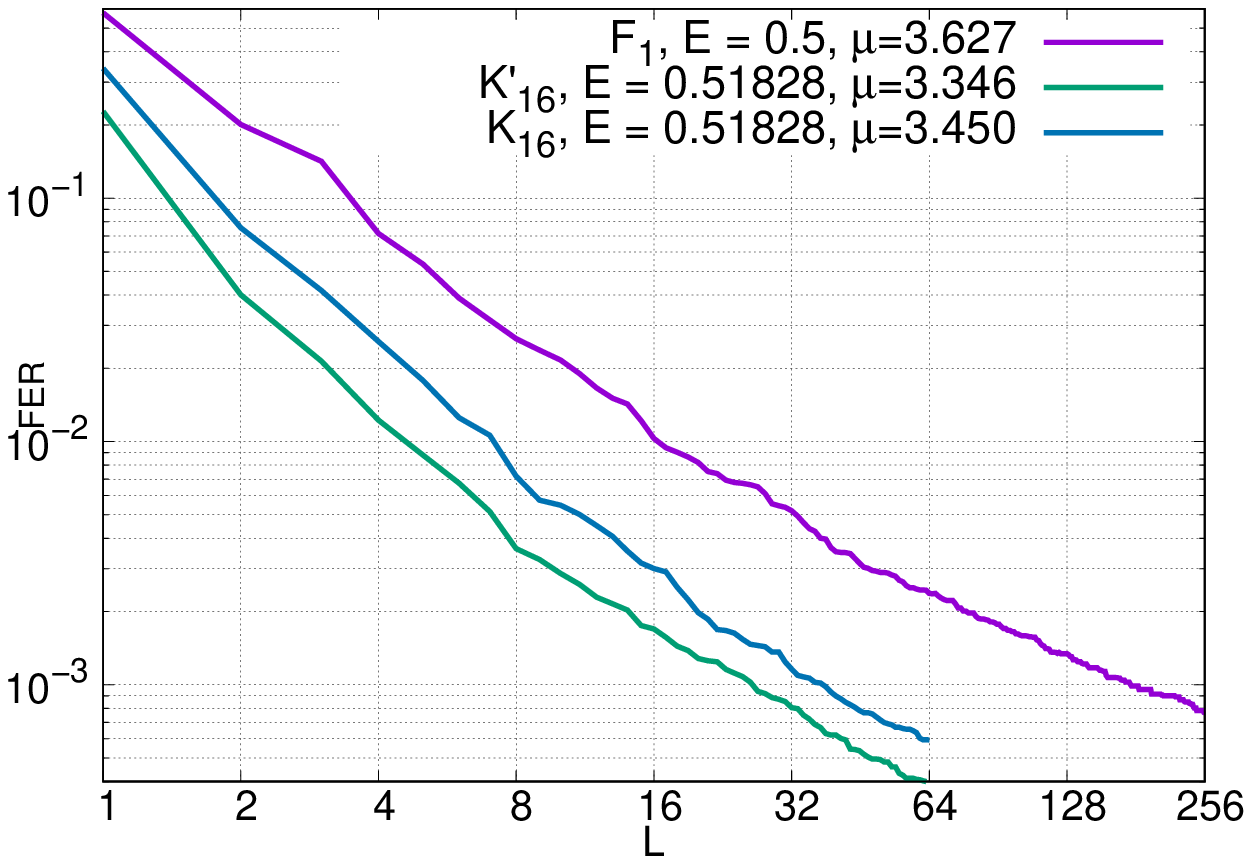}
\caption{Performance, $K_{16}, K'_{16}$}
\label{fErrorList}
\end{subfigure}
\begin{subfigure}[b]{0.2415\textwidth}
\includegraphics[width=1\linewidth]{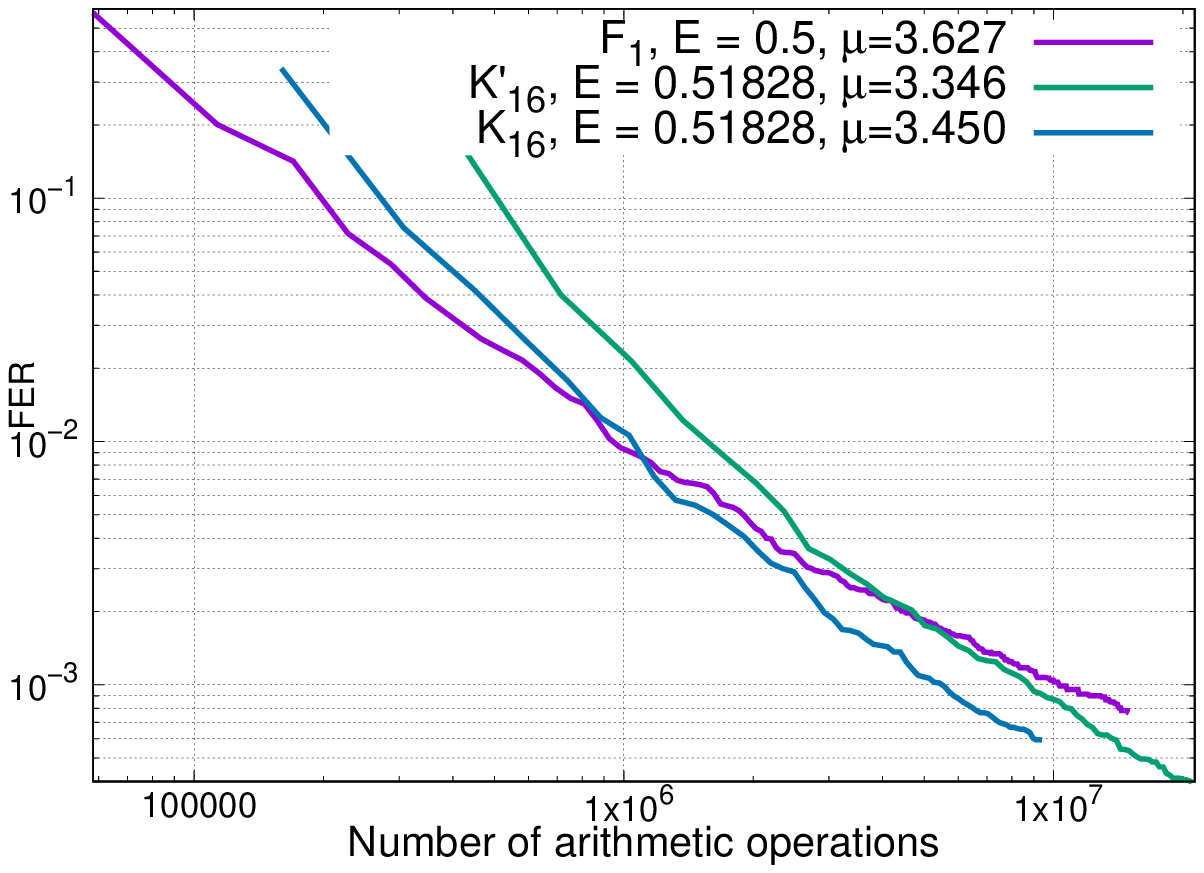}
\caption{Complexity, $K_{16}, K'_{16}$}
\label{fErrorCompl}
\end{subfigure}
\begin{subfigure}[b]{0.2415\textwidth}
\includegraphics[width=\linewidth]{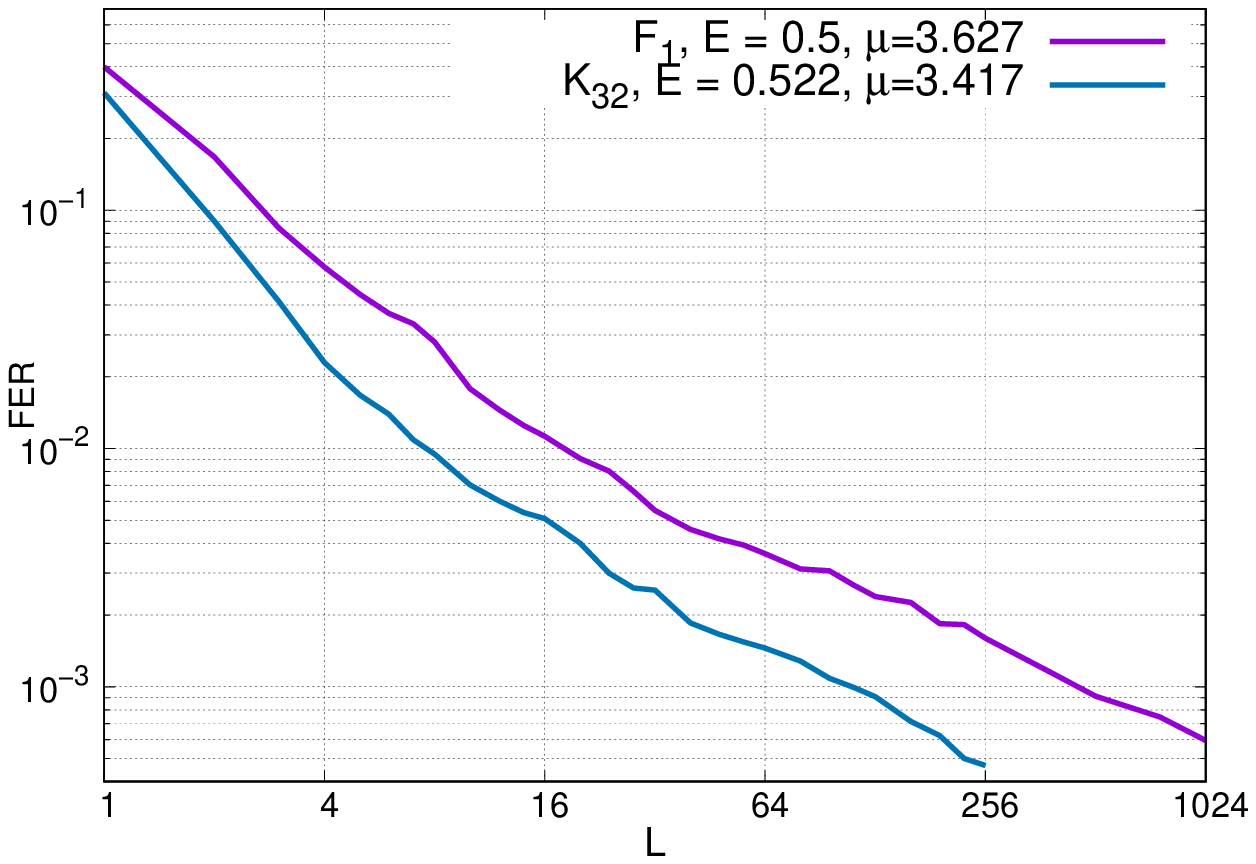}
\caption{Performance, $K_{32}$}
\label{fErrorList1024}
\end{subfigure}
\begin{subfigure}[b]{0.2415\textwidth}
\includegraphics[width=1\linewidth]{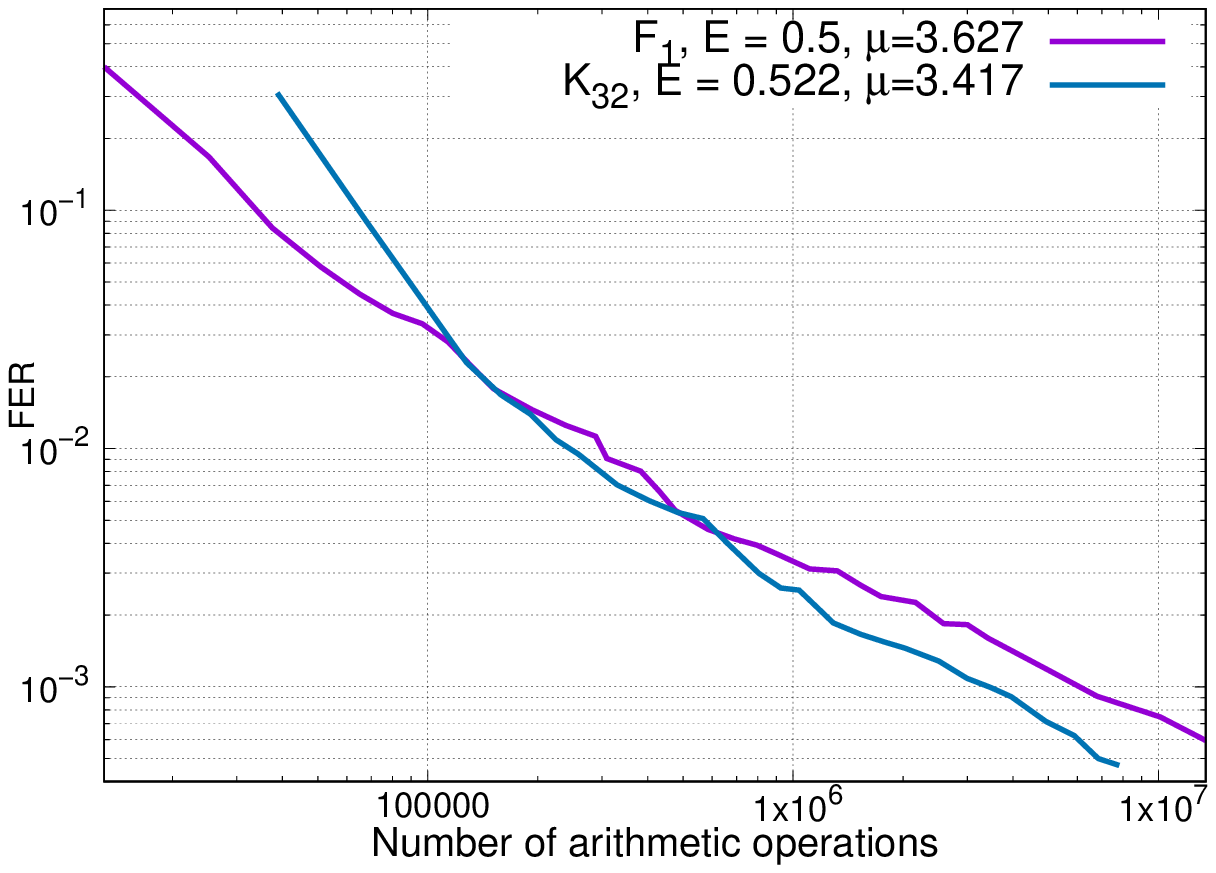}
\caption{Complexity, $K_{32}$}
\label{fErrorCompl1024}
\end{subfigure}
\caption{SCL decoding of polar subcodes with large kernels}
\end{figure}

Fig. \ref{fErrorList} presents the simulation results for $(4096,2048)$ PSCs with different kernels under SCL with varied $L$ at $E_b/N_0=1.25$ dB. It can be seen that the kernels with rate of polarization $0.51828$ require significantly lower list size $L$ to achieve the same performance as the code with $F_1$ kernel, which has the rate of polarization 0.5. Moreover, this gap grows with $L$. This is due to the improved rate of polarization, which results in a smaller number of unfrozen imperfectly polarized bit subchannels. The size of the list needed to correct possible errors in these subchannels grows exponentially with their number (at least for the genie-aided decoder considered in \cite{mondelli2015scaling}). On the other hand,   lower scaling exponent enables better performance with the same list size $L$, but the slope of the curve remains the same for both kernels $K'_{16},K_{16}$.

Fig. \ref{fErrorList1024} presents the simulation results for $(1024,512)$ PSC with different kernels under SCL with varied $L$ at $E_b/N_0=1.5$ dB. Similarly to the case of $16 \times 16$ kernels, $32 \times 32$ kernel $K_{32}$ with improved polarization properties ($E(K_{32}) = 0.521936, 
\mu(K_{32}) = 3.417$) provides significant performance gain compared with polar subcode with $F_1$. 

Fig. \ref{fErrorCompl} presents the performance of SCL\ decoding of $(4096,2048)$ PSCs with $K'_{16}$ and $K_{16}$ kernels in terms of the decoding complexity. 
  Observe that the PSC based on kernel $K_{16}$ can provide better performance with the same decoding complexity for FER $\leq 7\cdot 10^{-3}$ ($L \geq 8$ for $K_{16}$ and $L \geq 22$ for $K'_{16}$). This is due to the higher slope of the corresponding curve  in Fig. \ref{fErrorList}, which eventually enables one to compensate the relatively high complexity of the LLR computation algorithm presented in Section \ref{ssDescr16_2}.

Unfortunately, $K'_{16}$ kernel with lower scaling exponent has greater processing complexity than $K_{16}$, so that its curve intersects the one for  the $F_1$ kernel only at FER$=2\cdot 10^{-3}$.

Fig. \ref{fErrorCompl1024} presents the results of SCL\ decoding of $(1024, 512)$ PSC with $K_{32}$ in terms of the  decoding complexity. Similarly to $K_{16}$,  the PSC with $K_{32}$ can provide better performance with the same decoding complexity. Starting from FER $\leq 2 \cdot 10^{-2}$ PSC with $K_{32}$   ($L \geq 4$) maintain approximately the same performance-complexity tradeoff as PSC with $F_1$ ($L \geq 9$). Observe that $K_{32}$  enables one to reduce required list size by more than two times.  Starting with FER $\leq 4 \cdot 10^{-3}$ PSC with $K_{32}$ ($L \geq 20$) provides better performance with the same decoding complexity compared with PSC with $F_1$ ($L \geq 48$).

\subsection{Comparison with other large kernels}

\begin{figure}
\centering
\begin{subfigure}[b]{0.49\textwidth}
\includegraphics[width=\linewidth]{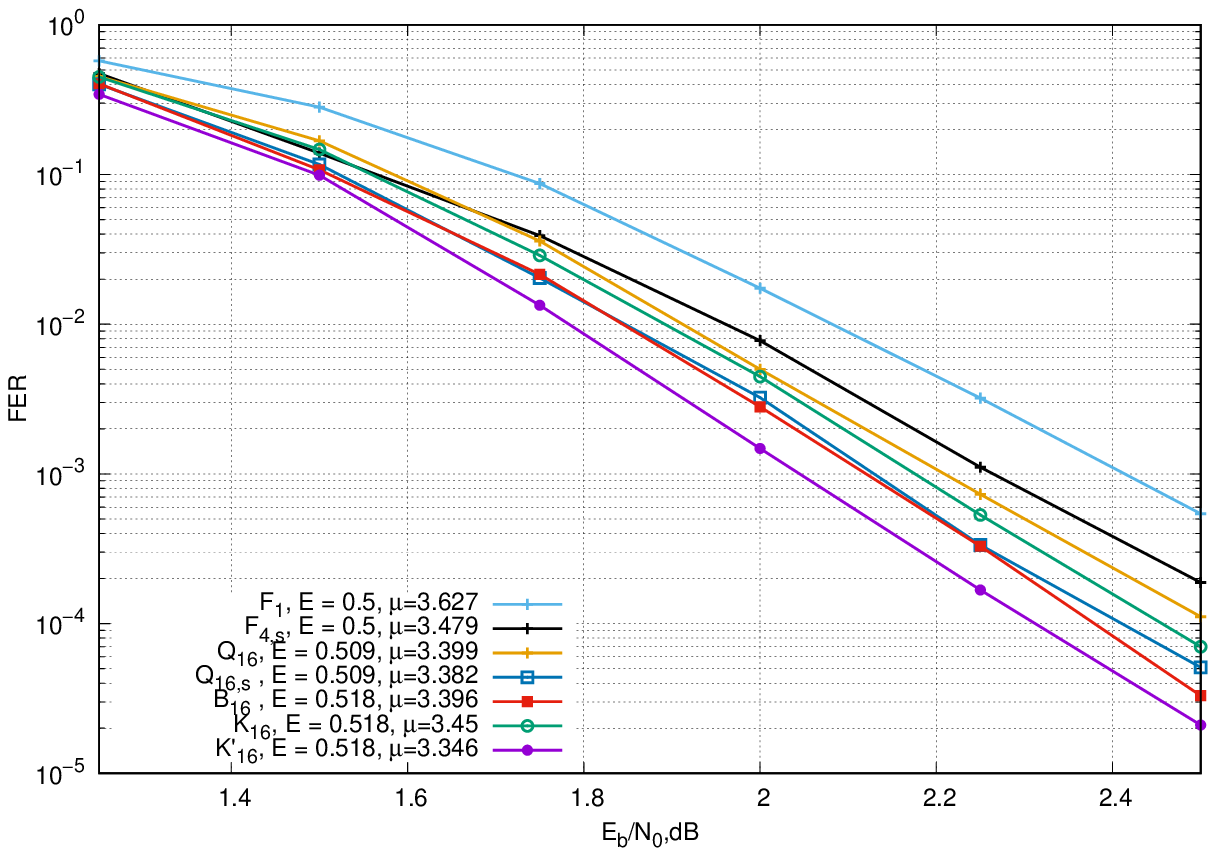}
\caption{$(4096, 2048)$ codes, $16\times 16$ kernels }
\label{f4096_2048_kernels}
\end{subfigure}
\begin{subfigure}[b]{0.49\textwidth}
\includegraphics[width=1\linewidth]{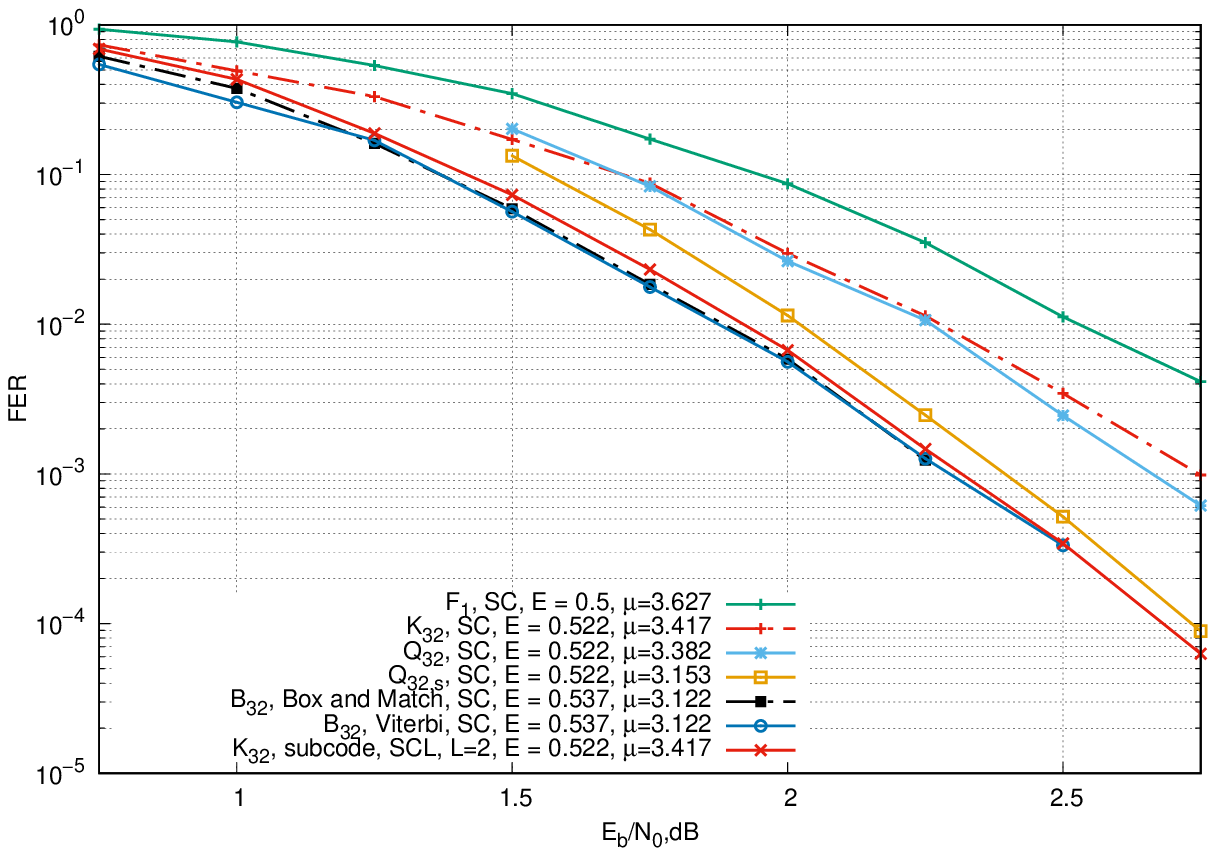}
\caption{$(1024, 512)$ codes, $32\times 32$ kernels }
\label{f1024_512_kernels}
\end{subfigure}
\caption{Performance of polar codes with various polarization kernels}
\end{figure}

\begin{table}[ht]
\centering
\caption{SC decoding complexity of (4096,2048) polar codes with various kernels}
\scalebox{1}{
\begin{tabular}{|c|c|c|c|c|c|}
\hline
$F_{4,s}$        & $Q_{16}$ & $K_{16}$         & $Q_{16,s}$  & $B_{16}$  & $K'_{16}$        \\ \hline
$1.7 \cdot 10^5$ & $2.1 \cdot 10^5$                      & $1.4 \cdot 10^5$ & $3.9 \cdot 10^5$                      & $1.3 \cdot 10^6$                      & $3.4 \cdot 10^5$ \\ \hline
\end{tabular}}
\label{f4096_2048_kernels_tab}
\end{table}
\begin{table}[ht]
\centering
\small
\caption{Decoding complexity of (1024,512) polar codes with various kernels}
\scalebox{1}{
\hspace{-2mm}
\begin{tabular}{|c|c|c|c|c|c|}
\hline
\multicolumn{5}{|c|}{SC}                                                                                                                                                                                                                              & SCL, L=2         \\ \hline
$K_{32}$         & $Q_{32}$        & $Q_{32,s}$       & \begin{tabular}[c]{@{}l@{}}$B_{32}$, \\ approx. \cite{miloslavskaya2014sequentialBCH}\end{tabular} & \begin{tabular}[c]{@{}l@{}}$B_{32}$, \\ Viterbi \end{tabular} & $K_{32}$         \\ \hline
$3.6 \cdot 10^4$ & $6.2 \cdot 10^4$ & $1.1 \cdot 10^6$ & $\approx 1.5 \cdot 10^5$                                                                           & $1.9 \cdot 10^7$                                                                          & $6.9 \cdot 10^4$ \\ \hline
\end{tabular}}
\label{f1024_512_kernels_tab}
\end{table}

In this section we compare the performance and complexity of the recently proposed kernels with kernels $K'_{16},K_{16}$ and $K_{32}$. We consider the convolutional polar kernels $Q_{16}$ and $Q_{32}$ \cite{morozov2020convolutional} processed by the algorithm introduced in \cite{morozov2020efficient}, as well as sorted convolutional polar kernels \cite{morozov2020convolutional} $Q_{16,s}$ and $Q_{32,s}$. Authors in \cite{moskovskaya2020design} minimized the complexity of trellis-based Viterbi processing algorithm for $16 \times 16$ and $32 \times 32$ BCH kernels $B_{16}$ and $B_{32}$. We also include the results for the Box and Match based approximate processing algorithm \cite{miloslavskaya2014sequentialBCH} applied to $B_{32}$ and the results for sorted Arikan kernel $F_{4,s}$ \cite{buzaglo2017permuted} processed by the WP\ algorithm.

Fig. \ref{f4096_2048_kernels} compares the performance of $(4096, 2048)$ polar codes with different kernels, Table \ref{f4096_2048_kernels_tab} depicts the decoding complexity of these codes. It can be observed that polar code with $K_{16}$  outperforms polar codes with $Q_{16}$ and $F_{4,s}$, while having lower processing complexity and better polarization properties (shown in the plot).  Polar code with $K'_{16}$ kernel also outperforms polar codes with $Q_{16,s}$ and $B_{16}$, while having lower processing complexity and better polarization properties.

Fig. \ref{f1024_512_kernels} compares the performance of $(1024, 512)$ polar codes with different kernels, while Table \ref{f1024_512_kernels_tab} depicts the decoding complexity of these codes. It can be seen that due to better polarization properties (shown in the plot) these kernels also provide better SC decoding performance. However, by employing SCL decoding of  with $L=2$, $K_{32}$ kernel can compete with $Q_{32,s}$  and $B_{32}$ in terms of performance, while having significantly lower arithmetic complexity. 

\section{Conclusions}
In this paper a reduced complexity decoding algorithm for polar codes with $2^t \times 2^t$ polarization kernels was proposed. Application of this approach to kernels of size $16$ and $32$ was considered. The algorithm computes the kernel input symbols LLRs via the ones for the Arikan kernel, and exploits the structure of recursive calculations induced by the Arikan kernel to identify and reuse the values of some common subexpressions. It was shown that in the case of SCL decoding with sufficiently large list size, the proposed approach results in lower decoding complexity compared with the case of polar (sub)codes with Arikan kernel with the same performance.

\section*{Acknowledgments}
We thank Fariba Abbasi Aghdam Meinagh for many comments and stimulating discussions. We also thank the editor and the reviewers for their beneficial comments.

\bibliographystyle{IEEEtran}

\begin{IEEEbiography}[{\includegraphics[angle=270,width=\textwidth]{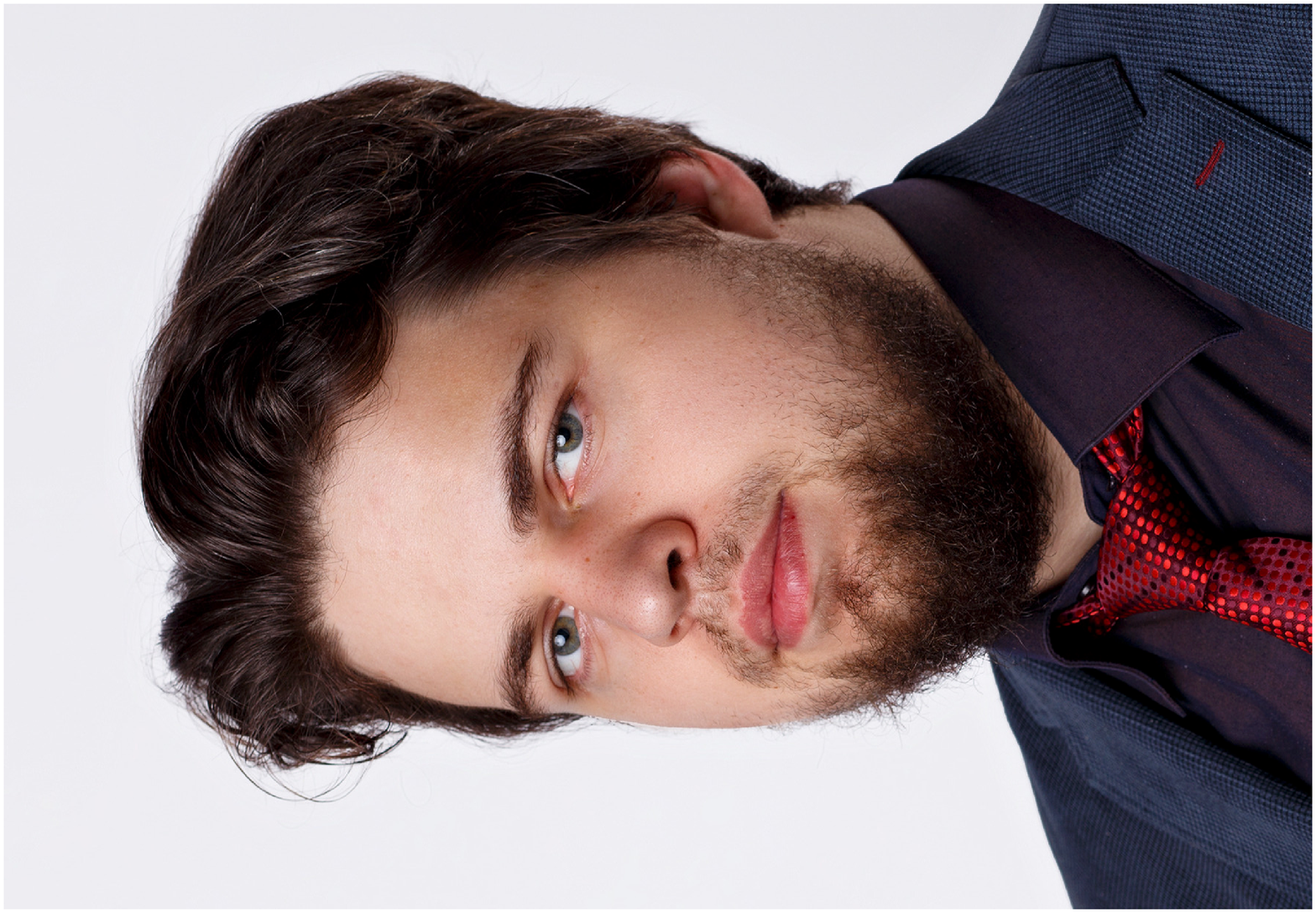}}]
{Grigorii Trofimiuk} (S'15)
was born in Boksitogorsk, Russia in 1994. He received 
the B.Sc. and M.Sc. degrees from St.Petersburg  Polytechnic
University in 2016 and 2018, respectively, all in computer science. He is currently working toward
the Ph.D. degree at the ITMO University in St.Petersburg, Russia. His research interests include coding theory and its applications in telecommunications.
\end{IEEEbiography}
\begin{IEEEbiography}[{\includegraphics[angle=270,width=\textwidth]{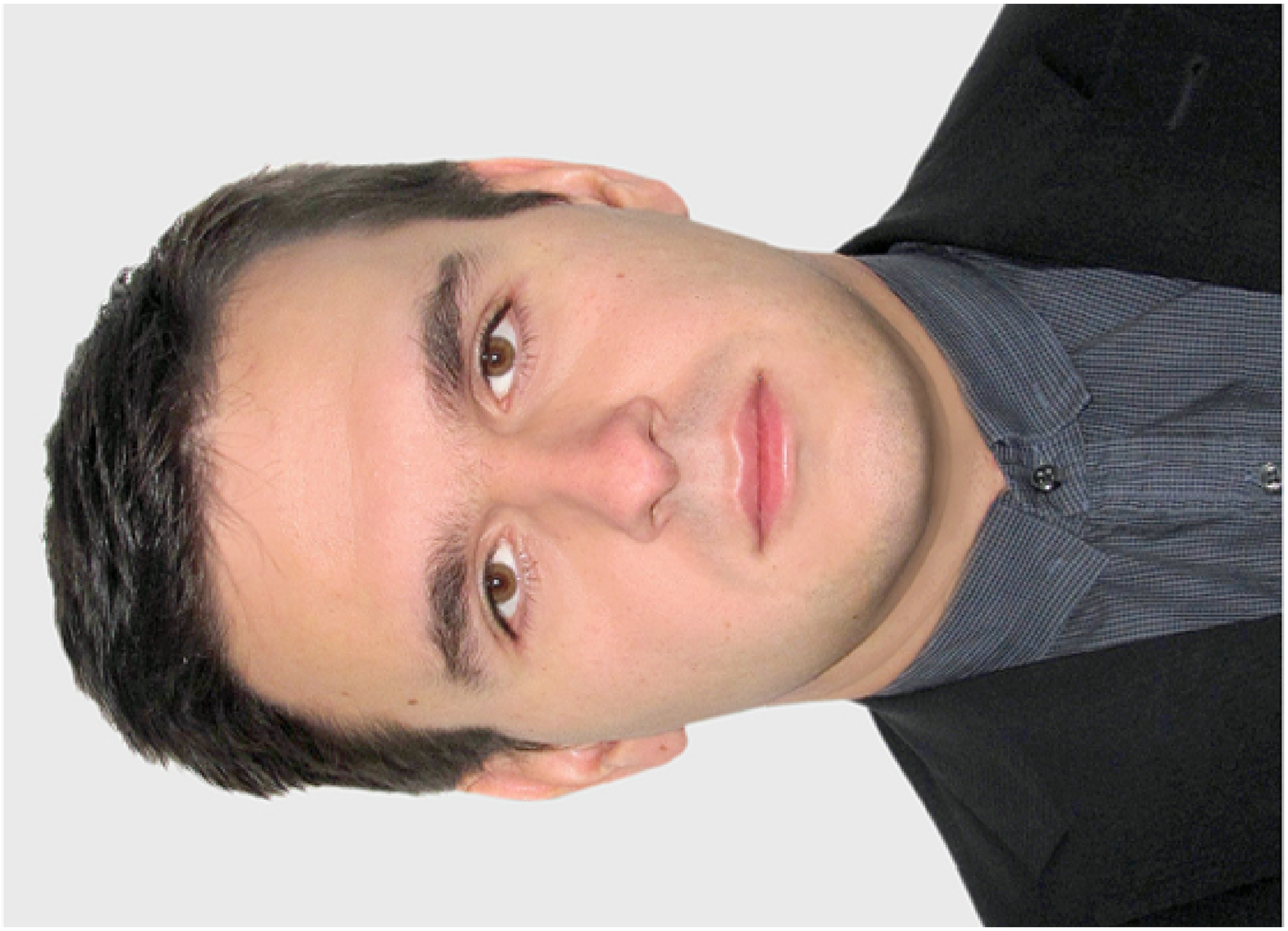}}]{Peter Trifonov} (S'02,M'05)
was born in St.Petersburg, USSR in 1980. He received 
the MSc and PhD (Candidate of Science) degrees from Saint Petersburg  Polytechnic
University in 2003 and 2005, and Dr.Sc degree from the Institute for Information Transmission Problems in 2018.  His research interests include coding theory and its applications in telecommunications and storage systems.  Currently he is a professor at the ITMO University in Saint Petersburg, Russia.
He is an editor at IEEE Transaction on Communications.
\end{IEEEbiography}

\vspace{-1mm}

\end{document}